\documentclass[aps,prb,longbibliography,numerical,preprint,showkeys]{revtex4-2}

\usepackage{epsfig,amsmath,amssymb,txfonts,hyperref,physics}
\usepackage{xcolor}

\newcommand{\Fig}[1]{Fig.~\ref{fig:#1}}

\newcommand{\Tab}[1]{Table~\ref{tab:#1}}

\newcommand{\Eqn}[1]{Eqn.~\ref{eqn:#1}}

\newlength{\wholefigwidth}
\newlength{\smallfigwidth}
\newlength{\halfsmallfigwidth}
\newlength{\figwidth}
\begin{document}

\title{Spin-polarized Energy Density Method from Spin-Density Functional Theory}

\author{Yang Dan}
\email{yangdan2@illinois.edu}
\author{Dallas R. Trinkle}
\email{dtrinkle@illinois.edu}
\affiliation{Department of Materials Science and Engineering, University of Illinois, Urbana-Champaign, Urbana, Illinois 61801, USA}

\date{\today}

\begin{abstract}
    The energy density method is generalized to include spin polarization with the full formalism derived based on spin-density functional theory, which aims at decomposing the total energy into well-defined atomic energies. The method involves two steps: (1) decomposing the total energy into spin-polarized energy density functions in real space, and (2) integrating these energy densities over chosen gauge-invariant volumes for uniquely defined atomic energies, whose summation over all the atoms restores the DFT total energy up to a constant difference. This method is numerically implemented into the Vienna \textit{ab initio} simulation package for the projector augmented-wave method, and is showcased with two applications. In the first application, we model the paramagnetic face-centered cubic Fe using spin special quasirandom structures; the spin energies are fit to spin cluster expansions and a deep neural network. In the second application, we calculate the atomic energy distributions of dilute magnetic semiconductor Ni-doped GaN with different dopant distances and spin configurations. This method extracts additional useful information for the study of magnetic systems with density functional theory.
\end{abstract}


\maketitle

\section{Introduction}

The total energy of a solid-state system is an extensive quantity, and can be viewed as the combined energies of its subsystems. For real materials which include crystalline defects, the subsystems are the individual defects along with the perfect bulk in the remaining space. Thus, the spatial distribution of the energy---not just the total energy---is an important quantity of interest. Chetty and Martin \cite{ChettyMartin1992Eden} decomposed the Kohn-Sham density functional theory (DFT) \cite{Hohenberg1964DFT,Kohn-Sham1965DFT} total energy into energy \textit{density} components at each position $\mathbf{r}$, enabling the calculation of a subsystem energy as integration over the subvolume of interest, compared to the conventional way of subtracting the total energies of two or more systems. This provides several benefits in practice: it saves computational effort for systems with multiple defects, as a single calculation is sufficient for evaluating all the defect energies; it reveals the energy distribution in the vicinity of a defect, which is more informative than the total energy as a single number and presents an intuitive picture of the system structure; and it calculates surface energies that cannot be studied via total energy differences due to geometric reasons \cite{ChettyMartin1992GaAs}.
However, the kinetic energy density and classical Coulomb components are non-unique by nature \cite{ChettyMartin1992Eden,ChettyMartin1991surfint}, where these different forms show gauge-dependence; the uniqueness of subsystem energies requires integrating over gauge-invariant subvolumes. Subsequent work by Yu, Trinkle and Martin \cite{Yu2011EDM,Yu2011Bader} applied Bader's theory of ``Atoms in Molecules'' \cite{Bader1990book,Bader1991paper} to find gauge-invariant subvolumes per atom in a solid-state system, which provides a general solution for the gauge-invariant subvolumes and well-defined atomic energies that only rely on electron density and potential solvable by DFT, removing any system-specific considerations in the analysis, such as crystalline symmetry \cite{ChettyMartin1991surfint,ChettyMartin1992GaAs,Rapcewicz1998}. The formulation of energy density in quantum systems, along with that of stress fields \cite{Filippetti2000, Shiihara2010} and in the analysis of chemical bonding, are further discussed in a recent paper\cite{Martin2025}.

While the energy density method has been well established so far as a generalized theoretical tool applicable to non-relativistic solid-state quantum systems, and proved its applications for various defective systems including surfaces/interfaces \cite{ChettyMartin1992GaAs,ChettyMartin1991surfint,Rapcewicz1998,Yu2011EDM,Yu2011TiO2,Lee2015EDM}, dislocations \cite{Dan2022,Dan2025} and point defects \cite{Ramprasad2002,Yu2011EDM}, the electron density remains the only degree of freedom subject to the variational principle for the ground state, and spin has not yet been taken into consideration. Spin is essential to studying magnetic systems, such as systems with transition metals and open-shell molecules, in which spins interact with each other or with an external magnetic field. This includes structural materials (e.g. steel and nickel alloys), spintronics and semiconductor quantum dots involving control of single localized spins \cite{Pulizzi2012}.

The rest of the paper is organized as follows. We first present the complete formalism of the spin-polarized energy densities, their gauge-invariant volumes, and the atomic energies. This is followed by two applications. In the first application, we study the magnetic exchange interactions in paramagnetic fcc Fe modeled using special quasirandom structure (SQS) \cite{Zunger1990SQS}, obtaining the magnetic configuration using spin-density functional theory (spin-DFT) \cite{Barth1972,Pant1972,Rajagopal1973,Vosko1980,Perdew1981,Jacob2012} and the atomic energies using our spin-EDM; their relationship is modeled combining Ginzberg-Landau-like self energies with exchange interactions by spin cluster expansion (SCE) \cite{Drautz2004SCE} or deep neural network (DNN). In the second application, we use spin-EDM  to determine the atomic energy landscape in DMS Ni-doped GaN, and changes with localized spins on Ni. These examples showcase the framework of analysis for atomic energies that provides additional information, insights and computational efficiency compared to standard spin-DFT.

\section{Methodology}

\subsection*{Energy density formalism from total energy in spin-DFT}

Spin-EDM is a generalization of EDM \cite{Yu2011EDM} to the case with spin polarization, and can be formulated in the framework of spin-DFT \cite{Barth1972,Pant1972,Rajagopal1973,Vosko1980,Perdew1981,Jacob2012}. Spin-DFT extends the spinless Hohenberg, Kohn and Sham DFT \cite{Hohenberg1964DFT,Kohn-Sham1965DFT} by incorporating the spin angular momentum as an extra degree of freedom, and considering its interaction with magnetic fields. The basic variable that uniquely determines the spin-dependent potential, the system total energy and the ground state is the spin density matrix, which is equivalently the scalar electron density $\rho(\mathbf{r})$ and the vectorized magnetization density $\mathbf{m}(\mathbf{r})$. Four assumptions underpin our formalism of spin-EDM: (1) the orbital magnetic moments or orbital currents of electrons are neglected; (2) local spin density approximation (LSDA) \cite{Barth1972,Perdew1981} and generalized gradient approximation (GGA) \cite{Perdew1996GGA,Becke1988} for the functional of exchange-correlation interactions, which assume that these interactions only depend on local electron and magnetization densities (for LSDA) as well as their spatial gradients (for GGA); (3) collinearity, which assumes that spin orientations are either parallel or anti-parallel to a global $z$-direction; and (4) spin-orbit coupling is neglected. The first two assumptions are inherited from spin-DFT; the other two assumptions are for simplicity while still covering a wide range of magnetism in real materials. With collinearity, the $x$ and $y$ components of $\mathbf{m}(\mathbf{r})$ are 0, and one can further simplify the vectorized $\mathbf{m}(\mathbf{r})$ to the scalar $m(\mathbf{r})$ as a basic variable. Same as in the spinless Kohn-Sham DFT, spin-DFT considers a reference system of non-interacting electrons with the same electron and magnetization densities as the real multi-electron system for constructing the orbitals $\psi_{i\sigma}(\mathbf{r})$.

Similar to the original expression in EDM \cite{Yu2011EDM}, we seek to derive an energy density function $e(\mathbf{r})$ in real space with consideration of spin polarization and with the form \footnote{The coupling term between magnetization $\mathbf{m}(\mathbf{r})$ and a magnetic field $\mathbf{B}(\mathbf{r})$ is not included with $e(\mathbf{r})$ because the influence from an external electromagnetic field is not a major consideration of our study, but we point out that its energy contribution $-\frac12 \int d\mathbf{r}\,\mathbf{m}(\mathbf{r}) \cdot \mathbf{B}(\mathbf{r})$ inherently leads to an energy density function of $-\frac12 \mathbf{m}(\mathbf{r}) \cdot \mathbf{B}(\mathbf{r})$}
\begin{equation}
	e(\mathbf{r}) = t(\mathbf{r}) + e_\text{CC}(\mathbf{r}) + e_\text{XC}(\mathbf{r}) + \sum_\mu E_\mu^\text{on-site} \delta(\mathbf{r} - \mathbf{R}_\mu),
		\label{eqn:e_EDM}
\end{equation}
where $t(\mathbf{r})$ is the kinetic energy density, $e_\text{CC}(\mathbf{r})$ is the classical Coulomb energy density, $e_\text{XC}(\mathbf{r})$ is the exchange-correlation energy density, and $\sum_\mu E_\mu^\text{on-site} \delta(\mathbf{r} - \mathbf{R}_\mu)$ is the on-site energy density that groups all the short-range energy terms, assigns them to ions, and is therefore represented as the sum of Dirac delta functions defined at the position $\mathbf{R}_\mu$ of each ion $\mu$. The exact expression of $E_\mu^\text{on-site}$ depends on the approach of pseudopotential, which will be covered in the following subsections and discussed specifically for projector-augmented wave (PAW) \cite{Blochl1994PAW,Kresse1999PAW,Hobbs2000spinPAW} method as well as norm-conserving pseudopotentials (NCPPs) \cite{Hamann1979NCPP,Troullier1991NCPP} and ultrasoft pseudopotentials (USPPs) \cite{Vanderbilt_USPP1990}. The integral of $e(\mathbf{r})$ over the entire volume should reproduce the system total energy $E_\text{tot}$, which, for our purpose, is written in the form of (in atomic units, $\hbar = m_e = e = 4\pi \epsilon_0 = 2 \mu_B = 1$)
\begin{equation}
    \begin{aligned}
        E_\text{tot} = &-\frac12 \sum_{\sigma=\uparrow,\downarrow} \sum_{n \mathbf{k}}  f_{n \mathbf{k}} \langle \tilde{\psi}_{n\mathbf{k}\sigma} | \nabla^2 | \tilde{\psi}_{n\mathbf{k}\sigma} \rangle + E_\text{XC} \left[ \rho^e(\mathbf{r})+\tilde{\rho}_c(\mathbf{r}), m^e(\mathbf{r}) \right] \\
        &+ \frac12 \int \rho^e(\mathbf{r}) V_\text{H}(\mathbf{r}) d\mathbf{r} + \int \sum_\mu V_\mu^\text{loc}(\mathbf{r}-\mathbf{R}_\mu) \rho^e(\mathbf{r}) d\mathbf{r} + \sum_{\mu<\nu} \frac{Z_\mu Z_\nu}{R_{\mu \nu}} + \sum_{\mu} E_\mu^\text{on-site}.
    \end{aligned}
    \label{eqn:E_tot}
\end{equation}
Here, $\tilde{\psi}_{n\mathbf{k}\sigma}$ is the pseudo-wavefunction for band $n$, wavevector $\mathbf{k}$ in the first Brillouin zone and the binary (up or down) spin $\sigma$, with $f_{n \mathbf{k}}$ being the occupation number; $\rho^e(\mathbf{r})$ is the valence electron density, which equals the pseudo-electron density for NCPPs, and the sum of the pseudo-electron density $\tilde{\rho}(\mathbf{r})$ and the compensation electron density $\hat{\rho}(\mathbf{r})$ for USPPs and the PAW method; $m^e(\mathbf{r})$ is the valence magnetization density. $\tilde{\rho}_c(\mathbf{r})$ is the pseudized frozen core electrons; $V_\text{H}(\mathbf{r})$ is the Hartree potential; and $V_\mu^\text{loc}(\mathbf{r}-\mathbf{R}_\mu)$ is the local part of the pseudopotential of ion $\mu$, and $Z_\mu$ is its ionic charge. The total energy $E_\text{tot}$ is represented as a combination of the non-interacting, pseudized electronic kinetic energy, the exchange-correlation energy, the Coulomb energies from electron-electron, ion-electron and ion-ion interactions, and short-range on-site energies assigned to each ion, respectively, from left to right in \Eqn{E_tot}, from which the exact form of the energy densities in \Eqn{e_EDM} is derived as follows.

\subsection{Kinetic energy density}

Kinetic energy density reflects the energy contribution from the first term in \Eqn{E_tot}. The kinetic energy density is not unique, and can be written in asymmetric or symmetric form as
	\begin{equation}
		\begin{aligned}
			t^{(a)}(\mathbf{r}) &= -\frac12 \sum_{n\mathbf{k}\sigma} f_{n\mathbf{k}}\tilde{\psi}_{n\mathbf{k}\sigma}^*(\mathbf{r}) \nabla^2 \tilde{\psi}_{n\mathbf{k}\sigma}(\mathbf{r}) \\
			t^{(s)}(\mathbf{r}) &= \frac12 \sum_{n\mathbf{k}\sigma} f_{n\mathbf{k}} \left| \nabla \tilde{\psi}_{n\mathbf{k}\sigma}(\mathbf{r})\right|^2
		\end{aligned}
		\label{eqn:ts_EDM_spin}
	\end{equation}
respectively, or as their linear combination. The difference between the asymmetric and symmetric forms yields the gauge-dependent term,
\begin{equation}
    t^{(a)}(\mathbf{r}) - t^{(s)}(\mathbf{r}) = -\frac14 \sum_{\sigma=\uparrow,\downarrow} \nabla^2 \tilde{\rho}_\sigma(\mathbf{r}) = -\frac14 \nabla^2 \tilde{\rho}(\mathbf{r}),
	\label{eqn:deltat_spin_EDM}
\end{equation}
which is proportional to the Laplacian of the corresponding smooth pseudo-electron densities. For spin-EDM, we choose the asymmetric form $t^{(a)}(\mathbf{r})$ to represent the kinetic energy densities, because with plane-wave basis, the Laplacians are well-defined (the pseudo-wavefunctions and pseudo-electron densities do not contain cusps) and are easy to calculate numerically via fast Fourier transforms (FFT) on regular grids between real and reciprocal space. 

\subsection{Classical Coulomb energy density}

Coulomb interactions are purely classical, so the Coulomb energy densities are by nature spin-independent, although it is possible to artificially partition the energy contributions to spin-polarized electron densities. Therefore, the formalism of classical Coulomb energy density $e_\text{CC}(\mathbf{r})$ in spinless EDM \cite{Yu2011EDM} is still applicable here in spin-EDM, 
\begin{equation}
	e_\text{CC}(\mathbf{r}) = e_\text{CC}^\text{(a),mod}(\mathbf{r}) = \left[ V^\text{loc}(\mathbf{r}) + \frac12 V_\text{H}(\mathbf{r}) + \frac12 V^\text{model}(\mathbf{r}) \right] \left[ \rho^e(\mathbf{r}) - \rho^\text{model}(\mathbf{r}) \right],
	\label{eqn:ecc_final_spin_EDM}
\end{equation}
where $V^\text{loc}(\mathbf{r}) = \sum_\mu V_\mu^\text{loc}(\mathbf{r}-\mathbf{R}_\mu)$ is the local part of the pseudopotential, $\rho^\text{model}(\mathbf{r})$ is the model charge density introduced to neutralize the rapidly varying local charge density arising from the local potential \cite{Yu2011EDM}, and $V^\text{model}(\mathbf{r})$ is the model potential arising from $\rho^\text{model}(\mathbf{r})$. The Coulomb energy density reflects the Coulomb energies from the third to the fifth terms in \Eqn{E_tot}, differing by a constant for each ion species if integrated \cite{Yu2011EDM}.

Similar to the case of kinetic energy density, the classical Coulomb energy density has two forms. One is the Maxwell form
\begin{equation}
	e_\text{CC}^\text{Maxwell}(\mathbf{r}) = \frac{1}{8\pi} \left| \nabla V^\text{tot}(\mathbf{r}) \right|^2.
	\label{eqn:ecc_Maxwell_EDM}
\end{equation}
The total classical Coulomb potential $V^\text{tot}(\mathbf{r}) = V_\text{H}(\mathbf{r})+V^\text{loc}(\mathbf{r})$ is the sum of Hartree potential $V_\text{H}(\mathbf{r})$ and the combined local potential $V^\text{loc}(\mathbf{r})$. The asymmetric form in \Eqn{ecc_final_spin_EDM} can constructed with a gauge transformation,
\begin{equation}
	e_\text{CC}^{(a)}(\mathbf{r}) = \frac12 V^\text{tot}(\mathbf{r})\rho^\text{tot}(\mathbf{r}) =  -\frac{1}{8\pi}V^\text{tot}(\mathbf{r}) \nabla^2 V^\text{tot}(\mathbf{r}),
	\label{eqn:ecca_spin_EDM}
\end{equation}
with
\begin{equation}
	\rho^\text{tot}(\mathbf{r}) = \rho^e(\mathbf{r})+\rho^\text{loc}(\mathbf{r}) = -\frac{1}{4\pi} \nabla^2 V^\text{tot}(\mathbf{r})
	\label{eqn:rhotot_Ecc_EDM}
\end{equation}
being the total electron density. The difference of the two forms gives the gauge-dependent term
\begin{equation}
	e_\text{CC}^{(a)}(\mathbf{r}) - e_\text{CC}^\text{Maxwell}(\mathbf{r}) = -\frac{1}{8\pi} \nabla \cdot \left[ V^\text{tot}(\mathbf{r}) \nabla V^\text{tot}(\mathbf{r}) \right]
	\label{eqn:deltaecc_EDM}
\end{equation}
which is the divergence of a function dependent on the total classical Coulomb potential $V^\text{tot}(\mathbf{r})$.

\subsection{Exchange-correlation energy density}

The exchange-correlation energy density $e_\text{XC}(\mathbf{r})$ is explicit under the assumption of LSDA, which makes the per-electron exchange-correlation energy functional local to $\mathbf{r}$. Define
	\begin{equation}
		\rho^{ec}_\sigma(\mathbf{r}) = \rho^e_\sigma(\mathbf{r}) + \tilde{\rho}_{c\sigma}(\mathbf{r}),
		\label{eqn:myrho_ec_EDM}
	\end{equation}
where $\rho^e_\sigma(\mathbf{r})$ is the pseudo-electron density for NCPPs or the valence electron density (sum of pseudo-electron density $\tilde{\rho}_\sigma(\mathbf{r})$ and compensation electron density $\hat{\rho}_\sigma(\mathbf{r})$) for USPPs and PAW, and $\tilde{\rho}_{c\sigma}(\mathbf{r})$ is the pseudized frozen core electron density. The frozen core is usually chosen to be full-shell so the core electron density can be seen as unpolarized, $\tilde{\rho}_{c\uparrow}=\tilde{\rho}_{c\downarrow}=\frac12 \tilde{\rho}_c$. The second term $E_\text{XC}$ in \Eqn{E_tot} is therefore written as
\begin{equation}
	E_\text{XC} \left[ \rho^{ec}_\uparrow(\mathbf{r}), \rho^{ec}_\downarrow(\mathbf{r}) \right] = \int d\mathbf{r} \,e_{\text{XC}}(\mathbf{r}),
	\label{eqn:EXC_spinEDM}
\end{equation}
with the exchange-correlation energy density $e_\text{XC}(\mathbf{r})$ defined as
\begin{equation}
    \begin{aligned}
        e_\text{XC}(\mathbf{r}) &= \sum_{\sigma=\uparrow,\downarrow} e_{\text{XC},\sigma}(\mathbf{r}) \\
        e_{\text{XC},\sigma}(\mathbf{r}) &= \rho^{ec}_\sigma(\mathbf{r}) \varepsilon_\text{XC} \left[ \rho_\uparrow^{ec}(\mathbf{r}), \rho_\downarrow^{ec}(\mathbf{r}), |\nabla \rho_\uparrow^{ec}(\mathbf{r})|, |\nabla \rho_\downarrow^{ec}(\mathbf{r})|, \cdots \right],
    \end{aligned}
	\label{eqn:eXC_spinEDM_LSDA}
\end{equation}
where $\varepsilon_\text{XC}$ is the per-electron exchange correlation energy, which is a functional of the spin-polarized electron densities for LSDA, and additionally their gradients or higher-order terms for GGA or further corrections.

An alternative representation uses the total electron and magnetization densities instead of the spin-polarized electron densities, which is commonly implemented in many software packages. The total electron density $\rho^{ec}(\mathbf{r})$ and magnetization density $m^{ec}(\mathbf{r})$ are
\begin{equation}
	\begin{aligned}
		\rho^{ec}(\mathbf{r}) &= \sum_{\sigma=\uparrow,\downarrow} \left[\rho_\sigma^e(\mathbf{r}) + \tilde{\rho}_{c\sigma}(\mathbf{r}) \right] = \rho^e(\mathbf{r}) + \tilde{\rho}_c(\mathbf{r}) \\
		m^{ec}(\mathbf{r}) &= \left[ \rho_\uparrow^e(\mathbf{r})-\rho_\downarrow^e(\mathbf{r}) \right] + \left[ \tilde{\rho}_{c\uparrow}(\mathbf{r})-\tilde{\rho}_{c\downarrow}(\mathbf{r}) \right] = \rho_\uparrow^e(\mathbf{r})-\rho_\downarrow^e(\mathbf{r}) = m^e(\mathbf{r}),
	\end{aligned}
\end{equation}
and the spin-polarized and the total exchange-correlation energy densities can be written as
\begin{equation}
	\begin{aligned}
		e_{\text{XC},\sigma}(\mathbf{r}) &= [\rho^e_\sigma(\mathbf{r})+\frac12 \tilde{\rho}_c(\mathbf{r})] \epsilon_\text{XC} \left[ \rho^e(\mathbf{r}) + \tilde{\rho}_c(\mathbf{r}), m^e(\mathbf{r}), \left| \nabla \rho^e(\mathbf{r}) + \nabla \tilde{\rho}_c(\mathbf{r}) \right|, |\nabla m^e(\mathbf{r})|, \cdots \right] \\
		e_\text{XC}(\mathbf{r}) &= \sum_{\sigma=\uparrow,\downarrow} e_{\text{XC},\sigma}(\mathbf{r}) = [\rho^e(\mathbf{r})+ \tilde{\rho}_c(\mathbf{r})] \epsilon_\text{XC} \left[ \rho^e(\mathbf{r}) + \tilde{\rho}_c(\mathbf{r}), m^e(\mathbf{r}), \left| \nabla \rho^e(\mathbf{r}) + \nabla \tilde{\rho}_c(\mathbf{r}) \right|, |\nabla m^e(\mathbf{r})|, \cdots \right]
	\end{aligned}
	\label{eqn:eXC_spinEDM_final}
\end{equation}

\subsection{On-site energy density}

We first present the formulation of on-site energy density based on the PAW method \cite{Blochl1994PAW,Kresse1999PAW,Hobbs2000spinPAW}. For magnetic systems, PAW is known to generally outperform NCPPs and USPPs in accuracy, especially for systems with localized \textit{d} or \textit{f} electrons. This is due to its ability to reconstruct the all-electron densities from smooth pseudo-wavefunctions, which provides more reliable magnetic moments and spin-state energetics, given that the exchange–correlation energy depends sensitively on the local spin density in the vicinity of atomic cores. Like other physical quantities in the PAW method, the total energy $E_\text{tot}$ is represented as the pseudized energy $\tilde{E}$ defined over the entire space, which is computationally evaluated on a regular plane-wave grid, plus the one-center correction inside the augmentation spheres near each ion $\mu$, which is the difference between all-electron and pseudized energies within the augmentation spheres $\Omega_\mu$ for each ion $\mu$ and evaluated on a radial grid \cite{Hobbs2000spinPAW}
\begin{equation}
		E_\text{tot} = \tilde{E} + \sum_\mu \left[E_\mu^1 - \tilde{E}_\mu^1\right]
		\label{eqn:Etot_PAW_EDM}
	\end{equation}
	with
	\begin{equation}
		\begin{aligned}
			\tilde{E} = &-\frac12 \sum_{\sigma=\uparrow,\downarrow} \sum_{n \mathbf{k}}  f_{n \mathbf{k}} \langle \tilde{\psi}_{n\mathbf{k}\sigma} | \nabla^2 | \tilde{\psi}_{n\mathbf{k}\sigma} \rangle + E_\text{XC} \left[ \tilde{\rho}_\uparrow+\hat{\rho}_\uparrow + \tilde{\rho}_{c \uparrow},  \tilde{\rho}_\downarrow+\hat{\rho}_\downarrow + \tilde{\rho}_{c \downarrow} \right] \\
			&+ \frac12 \int \rho^e(\mathbf{r}) V_\text{H}(\mathbf{r}) d\mathbf{r} + \int V_\text{H}\left[ \tilde{\rho}_{Zc} \right](\mathbf{r}) \rho^e(\mathbf{r}) d\mathbf{r} + \sum_{\mu<\nu} \frac{Z_\mu Z_\nu}{R_{\mu \nu}}\\
			\tilde{E}_\mu^1 = &\sum_{\sigma=\uparrow,\downarrow} \sum_{(i,j) \in \mu} -\frac12 \varrho_{ij}^\sigma\langle\tilde{\phi}_i | \nabla^2 | \tilde{\phi}_j\rangle + \overline{E_\text{XC}\left[\tilde{\rho}^1_\uparrow+\hat{\rho}_\uparrow+\tilde{\rho}_{c\uparrow}, \tilde{\rho}^1_\downarrow+\hat{\rho}_\downarrow+\tilde{\rho}_{c\downarrow} \right]} \\
			&+ \overline{E_\text{H}\left[\tilde{\rho}^1+\hat{\rho}\right]} + \int_{\Omega_r}V_\text{H}\left[\tilde{\rho}_{Zc}\right](\mathbf{r})\left[\tilde{\rho}^1(\mathbf{r})+\hat{\rho}(\mathbf{r})\right]d\mathbf{r} \\
			E_\mu^1 = & \sum_{\sigma=\uparrow,\downarrow} \sum_{(i,j) \in \mu} -\frac12 \varrho_{ij}^\sigma\langle\phi_i | \nabla^2 | \phi_j\rangle + \overline{E_\text{XC}\left[\rho^1_\uparrow+\rho_{c\uparrow}, \rho^1_\downarrow+\rho_{c\downarrow} \right]} \\
			&+ \overline{E_\text{H}\left[\rho^1\right]} + \int_{\Omega_r}V_\text{H}\left[\rho_{Zc}\right](\mathbf{r})\rho^1(\mathbf{r})d\mathbf{r}.
		\label{eqn:Es_PAW_EDM}
		\end{aligned}
	\end{equation}
Here, pseudized quantities are marked with tilde, compensation electron densities are marked with hat, and quantities with superscript 1 are one-center quantities; inside the augmentation sphere, $\varrho_{ij}^\sigma$ are occupancy matrices of augmentation orbitals $(i,j)$ and spin $\sigma$, and $\phi_i$ and $\tilde{\phi}_i$ are partial waves and pseudo partial waves of orbital $i$ that are dual to the projectors in PAW; the local potential is written as the Hartree potential of pseudized ionic charge density $\tilde{\rho}_{Z_c}$ ; and all other quantities share the same definition as in \Eqn{E_tot}. The on-site energy contains everything that is evaluated inside the augmentation sphere on radial grids, and is therefore defined as
\begin{equation}
	\begin{aligned}
		E^\text{on-site}(\mathbf{r}) &= \sum_\mu E_\mu^\text{on-site} \delta(\mathbf{r}-\mathbf{R}_\mu) \\
		E_\mu^\text{on-site} &= E_\mu^1 - \tilde{E}_\mu^1
	\end{aligned}
	\label{eqn:Eonsite_EDM}
\end{equation}
where $E_\mu^1$ and $\tilde{E}_\mu^1$ are one-center all-electron and pseudo energies defined in \Eqn{Es_PAW_EDM}. Both $E_\mu^1$ and $\tilde{E}_\mu^1$ contain contributions from kinetic, exchange-correlation, and classical Coulomb energies of electron-electron and ion-electron interactions, which are evaluated on radial grids inside the augmentation spheres.

For NCPP and USPP, the on-site energies $E_\mu^\text{on-site}$ are contributions of the non-local part of the pseudopotential. Their expressions are
\begin{equation}
    E_{\mu}^{\mathrm{on-site}} = \sum_{n\mathbf{k}\sigma} \sum_{\ell} \int d\mathbf{r} \,\tilde{\psi}_{n\mathbf{k}\sigma}^*(\mathbf{r}) \,V_{\mu \ell}^{\mathrm{nl}}\!\left(|\mathbf{r} - \mathbf{R}_\mu|\right) \,\varrho_{\ell} \,\tilde{\psi}_{n\mathbf{k}\sigma}(\mathbf{r})
    \label{eqn:Eonsite_EDM_NCPP}
\end{equation}
for NCPP \cite{Troullier1991NCPP,Yu2011EDM}, and
\begin{equation}
    E_{\mu}^{\mathrm{on-site}} = \sum_{n\mathbf{k}\sigma} \int d\mathbf{r} \,\tilde{\psi}_{n\mathbf{k}\sigma}^*(\mathbf{r}) \left(\sum_{ij} D_{ij}^{\mathrm{ion}} \, |\beta_i\rangle \langle \beta_j|\right)\tilde{\psi}_{n\mathbf{k}\sigma}(\mathbf{r})
    \label{eqn:Eonsite_EDM_USPP}
\end{equation}
for USPP \cite{Vanderbilt_USPP1990,Yu2011EDM}.

\begin{table*}[htp]
		\centering
		\caption{Spin-polarized energy densities and method of integration for atomic energies.}
		
		\centerline{
		\begin{tabular}{ll}
			\hline \hline
			Energy density \\
			
			$e(\mathbf{r}) = t(\mathbf{r}) + e_\text{CC}(\mathbf{r}) + e_\text{XC}(\mathbf{r}) + \sum_\mu E_\mu^{\text{on-site}} \delta\left( \mathbf{r} - \mathbf{R}_\mu \right)$ & \Eqn{e_EDM} \\
            
			\hline
			
			(1) Kinetic energy density \\
			
			$t^{(a)}(\mathbf{r}) = -\frac12 \sum_{n\mathbf{k}\sigma} f_{n\mathbf{k}}\tilde{\psi}_{n\mathbf{k}\sigma}^*(\mathbf{r}) \nabla^2 \tilde{\psi}_{n\mathbf{k}\sigma}(\mathbf{r}) \quad t^{(s)}(\mathbf{r}) = \frac12 \sum_{n\mathbf{k}\sigma} f_{n\mathbf{k}} \left| \nabla \tilde{\psi}_{n\mathbf{k}\sigma}(\mathbf{r})\right|^2$ & \Eqn{ts_EDM_spin} \\
			
		    $t(\mathbf{r}) = t^{(a)}(\mathbf{r})$ & \\
			
			Gauge dependence: $t^{(a)} (\mathbf{r}) - t^{(s)} (\mathbf{r}) = - \frac{1}{4} \nabla^2 \tilde{\rho} (\mathbf{r})$ & \Eqn{deltat_spin_EDM} \\
			
			Bader volume $\Omega_\rho$: $\nabla \tilde{\rho}(\mathbf{r}) \cdot \hat{\mathbf{n}}(\mathbf{r}) = 0 \quad \mathbf{r} \in \partial \Omega_\rho$ & \Eqn{Omega_rho_Bader} \\
			
			Integrated kinetic energy $T = \int_{\Omega_\rho} t(\mathbf{r}) d\mathbf{r}$ \\
            
			\hline
			
			(2) Classical Coulomb energy density \\
			
			$e_\text{CC}^\text{Maxwell}(\mathbf{r}) = \frac{1}{8\pi} \left| \nabla V^\text{tot}(\mathbf{r}) \right|^2$ & \Eqn{ecc_Maxwell_EDM} \\
			
			$e_\text{CC}^{(a)}(\mathbf{r}) = \frac12 V^\text{tot}(\mathbf{r})\rho^\text{tot}(\mathbf{r}) =  -\frac{1}{8\pi}V^\text{tot}(\mathbf{r}) \nabla^2 V^\text{tot}(\mathbf{r})$ & \Eqn{ecca_spin_EDM} \\
			
			
			$e_\text{CC}(\mathbf{r}) = e_\text{CC}^\text{(a),mod}(\mathbf{r}) = \left[ V^\text{loc}(\mathbf{r}) + \frac12 V_\text{H}(\mathbf{r}) + \frac12 V^\text{model}(\mathbf{r}) \right] \left[ \rho^e(\mathbf{r}) - \rho^\text{model}(\mathbf{r}) \right]$ & \Eqn{ecc_final_spin_EDM} \\
			
			Gauge dependence: $e_\text{CC}^{(a)}(\mathbf{r}) - e_\text{CC}^\text{Maxwell}(\mathbf{r}) = -\frac{1}{8\pi} \nabla \cdot \left[ V^\text{tot}(\mathbf{r}) \nabla V^\text{tot}(\mathbf{r}) \right]$ & \Eqn{deltaecc_EDM} \\
			
			Charge-neutral volume $\Omega_V$: $\nabla V^\text{tot}(\mathbf{r}) \cdot \hat{\mathbf{n}}(\mathbf{r}) = 0 \quad \mathbf{r} \in \partial \Omega_V$ & \Eqn{Omega_V_Bader} \\

			Integrated classical Coulomb energy $E_{\text{CC}} = \int_{\Omega_V} e_\text{CC}(\mathbf{r}) d\mathbf{r}$ \\

			\hline
			
			(3) Exchange-correlation energy density \\
			$e_{\text{XC},\sigma}(\mathbf{r}) = [\rho^e_\sigma(\mathbf{r})+\frac12 \tilde{\rho}_c(\mathbf{r})] \epsilon_\text{XC} \left[ \rho^e(\mathbf{r}) + \tilde{\rho}_c(\mathbf{r}), m^e(\mathbf{r}), \left| \nabla \rho^e(\mathbf{r}) + \nabla \tilde{\rho}_c(\mathbf{r}) \right|, |\nabla m^e(\mathbf{r})|, \cdots \right]$ & \Eqn{eXC_spinEDM_final} \\
			
			$e_\text{XC}(\mathbf{r}) = \sum_{\sigma=\uparrow,\downarrow} e_{\text{XC},\sigma}(\mathbf{r}) = [\rho^e(\mathbf{r})+ \tilde{\rho}_c(\mathbf{r})] \epsilon_\text{XC} \left[ \rho^e(\mathbf{r}) + \tilde{\rho}_c(\mathbf{r}), m^e(\mathbf{r}), \left| \nabla \rho^e(\mathbf{r}) + \nabla \tilde{\rho}_c(\mathbf{r}) \right|, |\nabla m^e(\mathbf{r})|, \cdots \right]$ & \Eqn{eXC_spinEDM_final} \\
			
			Bader volume $\Omega_\rho$: $\nabla \tilde{\rho}(\mathbf{r}) \cdot \hat{\mathbf{n}}(\mathbf{r}) = 0 \quad \mathbf{r} \in \partial \Omega_\rho$ & \Eqn{Omega_rho_Bader} \\
			Integrated exchange-correlation energy $E_{\text{XC}} = \int_{\Omega_\rho} e_{\text{XC}}(\mathbf{r}) d\mathbf{r}$ \\

			\hline
			
			(4) On-site energies \\
			
			PAW: $E_\mu^\text{on-site} = E_\mu^1 - \tilde{E}_\mu^1$, where $E_\mu^1$ and $\tilde{E}_\mu^1$ are defined in \Eqn{Es_PAW_EDM} & \Eqn{Eonsite_EDM} \\

            NCPP: $E_{\mu}^{\mathrm{on-site}} = \sum_{n\mathbf{k}\sigma} \sum_{\ell} \int d\mathbf{r} \,\tilde{\psi}_{n\mathbf{k}\sigma}^*(\mathbf{r}) \,V_{\mu \ell}^{\mathrm{nl}}\!\left(|\mathbf{r} - \mathbf{R}_\mu|\right) \,\varrho_{\ell} \,\tilde{\psi}_{n\mathbf{k}\sigma}(\mathbf{r})$ & \Eqn{Eonsite_EDM_NCPP} \\
            USPP: $E_{\mu}^{\mathrm{on-site}} = \sum_{n\mathbf{k}\sigma} \int d\mathbf{r} \,\tilde{\psi}_{n\mathbf{k}\sigma}^*(\mathbf{r}) \left(\sum_{ij} D_{ij}^{\mathrm{ion}} \, |\beta_i\rangle \langle \beta_j|\right)\tilde{\psi}_{n\mathbf{k}\sigma}(\mathbf{r})$ & \Eqn{Eonsite_EDM_USPP} \\
			\hline
			Atomic energy: $E = T + E_\text{CC} + E_\text{XC} + E_\mu^\text{onsite}$ \\
			\hline \hline
		\end{tabular}
	    }
		\label{tab:spin-EDM}
	\end{table*}
    
\subsection{Gauge-invariant volumes for integration of energy densities}
Similar to the spinless EDM, two gauge-dependent terms emerge in spin-EDM: kinetic energy density and classical Coulomb energy density. The kinetic energy density has a gauge-dependent term proportional to the Laplacian of the total pseudo-electron density, $\nabla^2 \tilde{\rho}(\mathbf{r})$, as shown in \Eqn{deltat_spin_EDM}. From Gauss's law, the integral of $\nabla^2 \tilde{\rho}(\mathbf{r})$ over volume $\Omega_\rho$ equals the overall flux of $\nabla \tilde{\rho}(\mathbf{r})$ through the surface of the volume $\partial \Omega_\rho$
	\begin{equation}
		\int_{\Omega_\rho} \nabla^2 \tilde{\rho}(\mathbf{r}) d\mathbf{r} = \int_{\partial \Omega_\rho} \nabla \tilde{\rho}(\mathbf{r}) \cdot \hat{\mathbf{n}}(\mathbf{r})dS,
		\label{eqn:gausslaw_rho}
	\end{equation}
where $\hat{\mathbf{n}}(\mathbf{r})$ is the unit normal vector at the surface. Therefore, if the volume $\Omega_\rho$ has a surface $\partial\Omega_\rho$ where the flux $\nabla \tilde{\rho}(\mathbf{r})$ is perpendicular to the normal $\hat{\mathbf{n}}(\mathbf{r})$, \Eqn{gausslaw_rho} is then automatically 0. Mathematically, these Bader volumes are defined as
\begin{equation}
	\begin{aligned}
		&\Omega_\rho: \quad \nabla \tilde{\rho}(\mathbf{r}) \cdot \hat{\mathbf{n}}(\mathbf{r}) = 0 \quad \mathbf{r} \in \partial \Omega_\rho \\
		&\Omega = \bigcup_\mu \Omega_{\rho \mu}, \quad \Omega_{\rho\mu} \bigcap \Omega_{\rho\nu} = \emptyset, \quad \mu \neq \nu,
		\end{aligned}
	\label{eqn:Omega_rho_Bader}
\end{equation}
where $\Omega$ is the entire volume of the system, and $\Omega_{\rho \mu}$ and $\Omega_{\rho \nu}$ are Bader volumes associated with ions $\mu$ and $\nu$, respectively, provided the fact that each Bader volume contains a single maximum of electron density located at an ion \cite{Bader1990book,Bader1991paper,Yu2011Bader}.

Similarly, for the classical Coulomb energy density, Gauss's law gives the integral of its gauge-dependent term in \Eqn{deltaecc_EDM} over subvolume $\Omega_V$ as
	\begin{equation}
		\int_{\Omega_V} \nabla \cdot \left( V^\text{tot}(\mathbf{r}) \nabla V^\text{tot}(\mathbf{r}) \right) d\mathbf{r} = \int_{\partial \Omega_V} V^\text{tot}(\mathbf{r}) \nabla V^\text{tot}(\mathbf{r}) \cdot \hat{\mathbf{n}}(\mathbf{r})dS
		\label{eqn:gausslaw_V}
	\end{equation}
and it is possible to set the integral to 0 by defining subvolume $\Omega_V$ as the zero-flux (charge-neutral) volume in the gradient vector field of the total potential, or the electric field, such that
\begin{equation}
	\begin{aligned}
		&\Omega_V: \quad \nabla V^\text{tot}(\mathbf{r}) \cdot \hat{\mathbf{n}}(\mathbf{r}) = 0 \quad \mathbf{r} \in \partial \Omega_V \\
		&\Omega = \bigcup_\mu \Omega_{V\mu}, \quad \Omega_{V\mu} \bigcap \Omega_{V\nu} = \emptyset, \quad \mu \neq \nu.
	\end{aligned}
	\label{eqn:Omega_V_Bader}
\end{equation}

The exchange-correlation energy density is not gauge-dependent as a result of LSDA, so in principle it can be integrated within an arbitrary subvolume. To keep consistency with Bader's theory for atomic properties, the integral of the exchange-correlation energy density is done in Bader volumes.

In spin-EDM, gauge-invariant integrations are always done by summing up the spin-polarized energy density components before doing integration over the Bader volume $\Omega_\rho$ defined by the total electron density $\tilde{\rho}(\mathbf{r})$.
\begin{equation}
	\begin{aligned}
		E_\text{atomic} &= \int_{\Omega_\rho} \sum_\sigma e_\sigma(\mathbf{r}) d \mathbf{r} \quad \mathbf{r} \in \partial \Omega_\rho \\
	\end{aligned}
\end{equation}
Using the total electron density $\tilde{\rho}(\mathbf{r})$ to define Bader volume boundaries is consistent with Bader's quantum theory for atoms in molecules \cite{Bader1990book,Bader1991paper}, in which the zero-flux volumes of the total electron density partition space for integration of atomic energies.

\section{Applications}

\subsection{Magnetic Exchange Interactions in fcc Fe}

The interaction of spins in a material can be described via magnetic exchange interactions. Perturbations from the ground state produce spin waves \cite{Stancil2009SpinWaves} and novel applications such as magnetic storage \cite{Comstock2002} and spintronics \cite{Pulizzi2012}. A model spin-based Hamiltonian can parameterize the change in energy with changes in spin. However, the common approach of modeling magnetic exchange interactions from DFT, which is sometimes called ``total energy difference method'' \cite{Xie2017YIG,Wang2008TbMnO,Gao2013}, is usually expensive. This method tracks the change in total energy when spins are flipped on different sites, which quantitatively determines the Hamiltonian by mapping the energy changes onto specific interaction terms or through data fitting. This generally involves costly calculations of multiple magnetic configurations for their precise total energies, and the cost grows significantly as the Hamiltonian grows complex or if the systems are large.

\begin{figure*}[htp]
	\makebox[\textwidth][c]{%
		\includegraphics[width=\smallfigwidth]{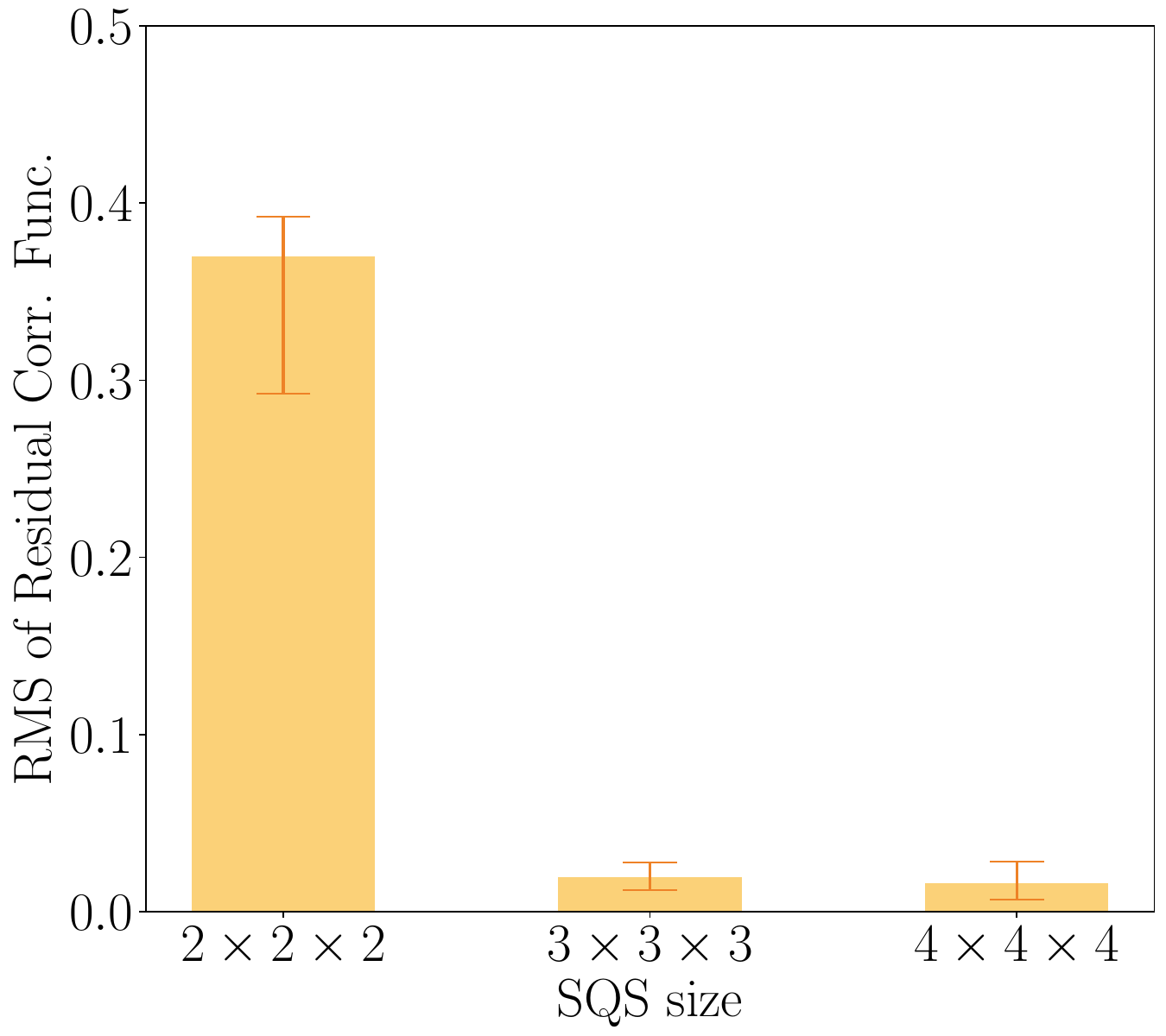}
	}
	\caption{Root mean square (RMS) of residual correlation functions of selected clusters (pair, triplet and quadruplet clusters with site distances up to the 10th nearest neighbor) in candidate SQS geometries of three different sizes (32-atom $2 \times 2 \times 2$, 108-atom $3 \times 3 \times 3$ and 256-atom $4 \times 4 \times 4$ supercells from the conventional unit cell of fcc Fe) that are generated using a Monte Carlo algorithm. The bar height indicates the average of RMS of residual correlation functions for all SQS's with the specified size, and the error bars indicate the minimal and the maximal values. Nine candidate SQS's are generated for each size. Good SQS candidates are indicated by overall small residual correlation functions. The $3 \times 3 \times 3$ SQS candidates show residual correlation functions comparable to $4 \times 4 \times 4$ SQS candidates, yet significantly smaller than the $2 \times 2 \times 2$ ones in general.}
	\label{fig:FCCFe_fom}
\end{figure*}

EDM has the advantage of observing an abundance of local energies influenced by local magnetic environments with only a few calculations, and therefore is potentially a more efficient way of studying the magnetism-energy interplay from first-principles. We apply EDM to study the magnetic exchange interactions in paramagnetic fcc Fe, which is a good system to observe for several reasons: (1) geometries are simple, with one-component Fe located at the lattice sites of a simple fcc structure, so the variation in atomic environments only comes from magnetism; (2) paramagnetism originates from randomness and disorder in local spin arrangements, which provides large variations in local magnetic environments; and (3) sophisticated ways have been proposed and applied to model paramagnetism in DFT supercells for fcc Fe or Fe alloys, such as by using special quasirandom structures (SQS) \cite{Zunger1990SQS,Abrikosov2016pm,Ponomareva2020FeMn}. We generate the initial spin configurations using a Monte Carlo algorithm for SQS \cite{VanDeWalle2013MCSQS}; for each structure, $10^6$ Monte Carlo steps are performed and the spin configuration with the optimal objective function \cite{VanDeWalle2013MCSQS}, evaluated on pair clusters with diameters up to the 10th nearest neighbor, triplet clusters up to the 5th nearest neighbor and quadruplet clusters up to the 3rd nearest neighbor, is selected. Here, the diameter of a cluster is defined as the maximum distance between any two sites in the cluster. This process has been applied to generate SQS's of three sizes: 32-atom $2 \times 2 \times 2$, 108-atom $3 \times 3 \times 3$ and 256-atom $4 \times 4 \times 4$ supercells, with each size having nine candidate structures. All candidate structures are evaluated for the root mean square (RMS) of residual correlation functions, with site distances up to the 10th nearest neighbor for pair, triplet and quadruplet clusters; the result reveals that the $3 \times 3 \times 3$ supercells have similar levels of residual correlation functions compared to the $4 \times 4 \times 4$ ones, but significantly lower than the $2 \times 2 \times 2$ ones, as shown in \Fig{FCCFe_fom}. Thus, the $3 \times 3 \times 3$ supercells best balance between the ability to accurately capture the disorder of spins and computational efficiency. We therefore pick three $3 \times 3 \times 3$ supercells with the lowest RMS of residual correlation functions for the DFT and EDM calculations that follow, which are referred to as SQS 1, 2 and 3, respectively.

We develop empirical models that predict EDM atomic energies from the magnetic configuration characterized by magnetic moments of atoms on the lattice sites, $\{ \mathbf{M}_j \}$. The EDM atomic energies are referenced with the energy of bulk fcc Fe atom with no spin polarization, so the atomic energies reflect only the magnetic contribution to the energies. For a system with spontaneous magnetization, the magnetic energy $E_i$ of site $i$ contains contributions from two parts: a self-spin term $E_i^\text{self}$ that only depends on the magnetic moment $\mathbf{M}_i$ of itself, and contribution $E_i^\text{e}$ from its exchange interactions with neighbors:
\begin{equation}
	E_i(\{ \mathbf{M}_j \}) = E_i^\text{self}(\mathbf{M}_i) + E_i^\text{e}(\{\mathbf{M}_j\}).
\end{equation}
The self-spin energy $E_i^\text{self}(\mathbf{M}_i)$ can be approximated by a Landau function, which is the power series of the magnitude of the magnetic moment; only even powers are included because of the symmetry with inverted $\mathbf{M}_i$. The simplest form of Landau function is
\begin{equation}
	E_i^\text{self}(\mathbf{M}_i) = -A \mathbf{M}_i^2 + B \mathbf{M}_i^4,
\end{equation}
where $A$ and $B$ are both positive numbers. The local minimum of $E_i^\text{self}(\mathbf{M}_i)$ leads to the solution of finite magnetic moment $|\mathbf{M}_i| = \sqrt{\frac{A}{2B}}$. The exchange interactions can be minimally modeled by an Ising model, where the longitudinal fluctuations of the magnetic moments are ignored. In an Ising model, the interaction energies are expressed as the linear combination of spin pairs, where the spins can be either $+1$ or $-1$. A more complex model is the Heisenberg model in which binary spins are replaced by magnetic moments to capture the longitudinal fluctuations, and transverse fluctuations if the directions of magnetic moments are arbitrary (non-collinear magnetism)
\begin{equation}
	E_i^\text{e, Heisenberg} = -\sum_\alpha J_\alpha \sum_{i, j \in \alpha} \mathbf{M}_i \cdot \mathbf{M}_j.
\end{equation}
For fcc Fe, it is important to consider the longitudinal fluctuation of magnetic moments as well as non-Heisenberg interactions involving more than two sites \cite{Singer2011}. Therefore, we apply a spin-cluster expansion (SCE) model \cite{Drautz2004SCE} to comprehensively take these factors into consideration.

The SCE model can be expressed as
\begin{equation}
	E_i^\text{e, SCE}(\{ \mathbf{M}_j \}) = -\sum_{\Omega_\alpha^i} J_\alpha \sum_{\delta \in \Omega_\alpha^i} \Gamma_\delta(\{ M_j \}),
	\label{eqn:SCE}
\end{equation}
where $\Omega_\alpha^i$ is the collection of all clusters containing site $i$ that are identical to cluster $\alpha$ by symmetry, $J_\alpha$ is the coupling constant for exchange interaction of cluster $\alpha$, which parameterizes the SCE model, and $\Gamma_\delta = \prod_{k \in \delta} M_k$ is the cluster function for cluster $\delta$, where $\mathbf{M}_k = M_k \mathbf{e}$ given that $\mathbf{e}$ is the unit vector that aligns with the common axis of the collinear spins. The Heisenberg model is a special case of the SCE model when only pair clusters are included. SCE model is a linear model of the interactions $\sum_{\delta \in \Omega_\alpha^i} \Gamma_\delta(\{ M_j \})$ and can be fitted with least squares method.

Alternatively, we develop a fully connected deep neural network (DNN) model to predict the exchange energy of each atom based on the magnetic environment in its vicinity. The architecture of the DNN model follows the design of a machine learning model for configurational energy used by Natarajan \textit{et al.} \cite{Natarajan2018}, which uses site-centric correlation functions based on the cluster expansion approach for input features. The site-centric correlation function at site $i$ is defined as
\begin{equation}
	\rho_\alpha^i = \sum_{\delta \in \Omega_\alpha^i} \Gamma_\delta(\{M_j\}),
\end{equation}
which has the same form as used in the SCE model. $\rho_\alpha^i$ is invariant to point group symmetry operations, so the crystal symmetry is automatically incorporated in the neural network. Each input node takes a correlation function $\rho^i_\alpha$ corresponding to a representative cluster $\alpha$, which is symmetrically different from the representative clusters in other input nodes; therefore, the number of input nodes equals the number of symmetrically unique clusters from the input data. Three fully-connected hidden layers are introduced after the input layer, with each followed by an activation layer with Leaky Rectified Linear Unit (ReLU) activation function. The output layer contains one node which outputs the prediction of the exchange energy $E_i^\text{e, DNN}$ at site $i$. Different from the linear SCE model, the DNN model considers both linear and nonlinear combinations of the input correlation functions; therefore, we use the DNN model to benchmark the limit of accuracy for the SCE models trained with the same data. The parameters in the Landau-DNN models, including the weights in the neural network and the coefficients $A$ and $B$ in the Landau function, are trained simultaneously to minimize the loss function, defined as the mean squared error (MSE) between the ground-truth EDM atomic energies and Landau-DNN model predictions. Adam optimizer is used to optimize the model parameters in training, with learning rate 0.001 and L2 regularization. The models are trained for 8000 epochs with 5-fold cross-validation. The number of nodes in the hidden layers are 6, 6 and 3, which are determined as hyperparameters that yield the lowest validation errors. Mini-batch training is enabled with batch size of 128.

Both the SCE and DNN models rely on the correlation function $\rho_\alpha^i$ of selected clusters, and we hereby apply a greedy algorithm for selecting the clusters with the greatest importance from a pool, described as follows. Starting from a Landau-SCE model with two parameters $A$ and $B$ for the Landau self-spin energy, and 0 parameters or clusters for the exchange energy, we check the Pearson's correlation
\begin{equation}
	r_\alpha = \sum_i \frac{\text{cov}(\Delta E_i, \rho_\alpha^i)}{\sigma_{\Delta E_i} \sigma_{\rho_\alpha^i}}
\end{equation}
between the energy residue of the current Landau-SCE model $\Delta E_i = E_i^\text{EDM} - (E_i^\text{self}+E_i^\text{e,SCE})$ and the candidate input feature of the correlation function $\rho_\alpha^i$ for cluster $\alpha$. We keep the cluster $\alpha^*$ that yields the greatest $|r_\alpha|$ among all the candidate clusters. Then we include cluster $\alpha^*$ in the SCE model, refit the energy residue $\Delta E_i$, and check its Pearson's correlation with the correlation functions of the clusters in the rest of the pool. This procedure is repeated until the required number of clusters are selected.

DFT calculations with spin-polarization are performed to relax the atom positions, followed by self-consistency calculations to determine the stable charge densities and magnetic configurations. The calculations are performed with Vienna \textit{Ab initio} Simulation Package (VASP) \cite{Kresse1993,Kresse1994,Kresse1996VASPa,Kresse1996VASPb,Kresse1999PAW} version 5.4.4. Spin-polarized EDM calculations are performed following the DFT calculations for the atomic energies in the three SQS's. The EDM calculations use the same parameters as DFT, except that the density of real-space grids are created with a uniform spacing of $0.064$ $\text{\AA}$ in each dimension, which is twice as dense as used in the DFT relaxations, to ensure that the gauge-dependent errors in EDM atomic energies fall below 1 meV/atom. In these calculations, a plane-wave basis is used with a cut-off energy 550 eV to converge both the total energy from DFT and atomic energies from EDM to less than 1 meV/atom. The energy tolerance for self-consistency calculations is $10^{-8}$ eV. The projector augmented-wave (PAW) pseudopotential \cite{Blochl1994PAW,Kresse1999PAW,Hobbs2000spinPAW} with Ar core and 8 valence electrons ($3d^74s^1$) is used, with Perdew-Burke-Ernzerhof (PBE) functional \cite{Perdew1996GGA} for the exchange-correlation energy. Methfessel-Paxton smearing \cite{Methfessel1989} with order 1 and an energy width of $0.25$ eV is applied. The k-point mesh grids are gamma-centered with uniform spacing of around $0.098$ $\text{\AA}^{-1}$ in each dimension in the k-space. The smearing and k-point mesh are chosen so that the density of states (DOS) of bulk fcc Fe near the Fermi level is in good agreement with the DOS calculated by the linear tetrahedron method with Bl\"{o}chl corrections \cite{Blochl1994Tetra} and with a dense k-point mesh, without spin polarization. The lattice parameter for the fcc Fe unit cell is $a_0 = 3.573$ $\text{\AA}$, which is determined through extrapolating the room temperature lattice parameters of Fe-C and Fe-N retained austenites to zero solute concentration \cite{Cheng1990}. The initial magnitude of magnetic moments of each atom is set to 3.0 $\mu_\text{B}$, with the sign determined from the SQS configurations. The atomic energy of Fe atoms from nonspin-polarized EDM calculation of a 4-atom fcc conventional unit cell is used as reference for the EDM calculations of the SQS's.

\begin{figure}[htb]
	\includegraphics[width=\wholefigwidth]{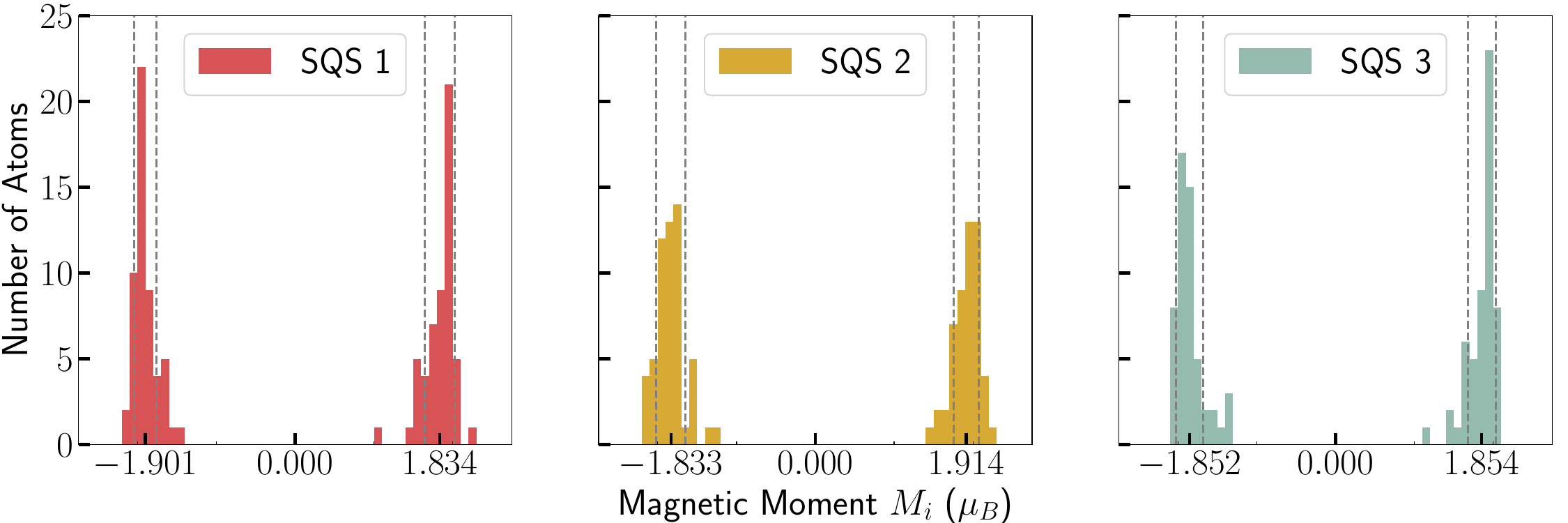} 
    \caption{Histogram of atomic magnetic moments $M_i$ in the three fcc paramagnetic Fe SQS's as determined by spin-DFT. $M_i$ are distributed around $\pm 1.85 \mu_\text{B}$ with the signs indicating the direction of spins. Each bin in the histogram has a width of $0.1 \mu_\text{B}$. The numbers mark the average value of $M_i$ for atoms with the same spin direction and the grey dashed lines mark the range of one standard deviation around the averaged $M_i$. The fluctuation of the magnitudes of $M_i$ indicates the influence of non-Ising spin interactions in the system.}
	\label{fig:sqsmagmom}
\end{figure}

The DFT calculations show that the magnetic moments in each SQS fall into two groups of opposite signs, corresponding to up and down spin directions, with narrow variations in magnitude, as shown in the histograms in \Fig{sqsmagmom}. The average magnitudes of the magnetic moments in each group are marked in the horizontal axes, with the range of one standard deviation from the average marked as grey dashed lines. Generally, the distributions of magnetic moments in the three SQS's show similarity with average magnitude falling in the range of $1.8 \sim 1.9\, \mu_\text{B}$ and standard deviation of $0.14 \sim 0.19\, \mu_\text{B}$; the average magnetic moments are close to other reported values: $2.08\, \mu_\text{B}$ calculated using DFT at 1400 K \cite{Zhang2011PMFe}, $1.89\, \mu_\text{B}$ calculated using dynamical mean-field theory (DMFT) \cite{Zhang2011PMFe, Leonov2011}, and $1.9\, \mu_\text{B}$ at 1200 K calculated using disordered local moment (DLM) models \cite{Ehteshami2017}. After relaxation of the Fe atoms and the electronic structures, none of the magnetic moments in our SQS cells have flipped from their initial arrangement. The magnitudes have changed, but the net magnetization of each supercell is below $4 \mu_\text{B}$ over the entire 108-atom supercells.

\begin{figure}[htb]
	\includegraphics[width=\figwidth]{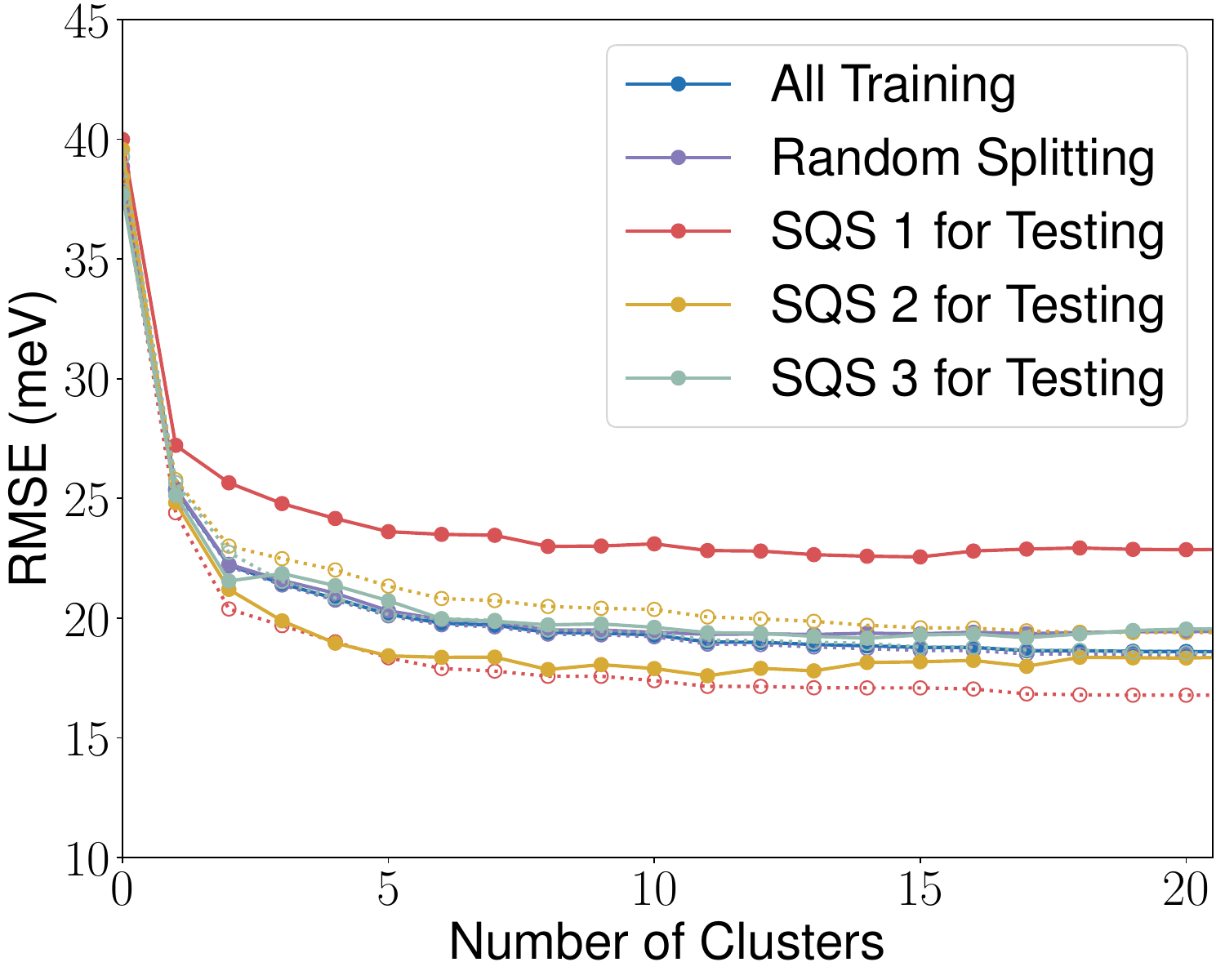} 
	\includegraphics[width=\figwidth]{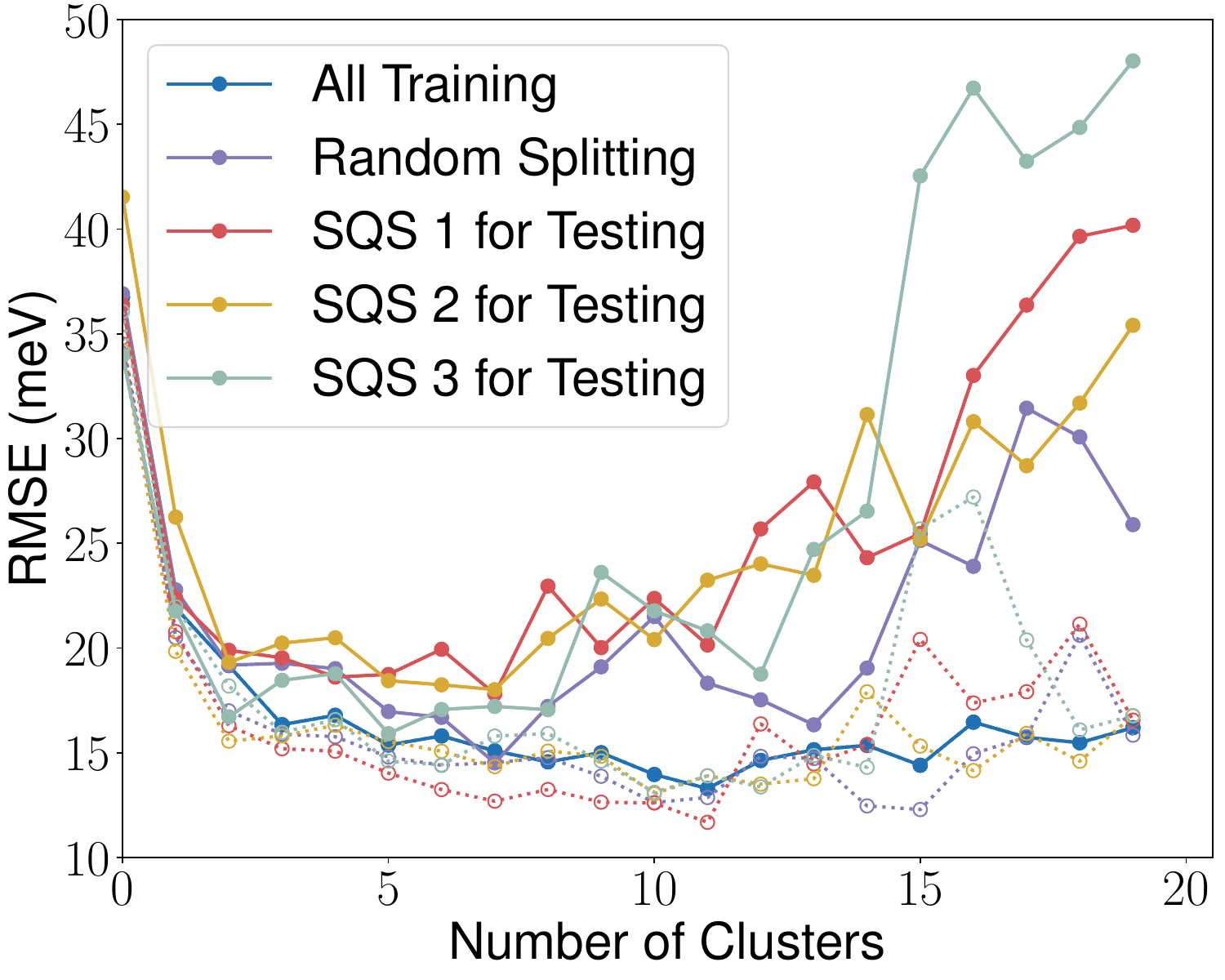} 
	\caption{Training (dashed lines) and testing (solid lines) root mean square error (RMSE) plots of the Landau-SCE models (left) and the Landau-DNN models (right) that use different numbers of most important clusters selected by the greedy algorithm. Five ways of training-testing data splitting are performed: use all the data available for training (all training), random 80\%-20\% splitting (random splitting), and SQS-based splittings in which each SQS is taken for testing while the other two SQS's are used for training. Both SCE and DNN models show that the minimal testing RMSEs can be achieved with the first 5 clusters selected by the greedy algorithm. The DNN models see rapidly increasing training errors when number of clusters exceeds 8, suggesting that over complex  models lead to overfitting with the current data.}
	\label{fig:rmses_fccFe}
\end{figure}

The Landau-SCE model is first applied with the greedy algorithm for cluster selection, with the pool of candidate clusters containing pair clusters of diameters from the 1st nearest neighbor up to the 7th nearest neighbor, and quadruplet clusters with diameters up to the 7th nearest neighbor. Only even numbers of sites are considered for the clusters to ensure that the cluster functions $\Gamma_\alpha(\{M_j\}) = \prod_{k \in \delta} M_k$ obey inversion symmetry. Pair clusters from the first to the fourth nearest neighbors are selected by the greedy algorithm as the top 4 most important clusters, followed by a tetrahedron quadruplet cluster with the sites being first nearest neighbors of each other and occupying the corners of the tetrahedron. Pair clusters containing 5th, 6th and 7th neighbors appear as the 13th, 12th and 8th most important clusters in the selection. The left figure in \Fig{rmses_fccFe} shows the training and testing root mean square errors (RMSEs) of the Landau-SCE models trained and evaluated using the top $N$ clusters selected, plotted as function of $N$. Five different ways of data splitting are used: use all the magnetization and EDM atomic energy data from the three SQS's as the training set (all training), randomly pick $80\%$ of the data for training and use the remaining $20\%$ for testing (random splitting), and SQS-based splitting in which data from one SQS is used as the testing dataset while two other SQS's are used for training. Result shows that the RMSEs decrease rapidly with the first 5 clusters, but change slowly or almost stay constant with more clusters included, suggesting that the first 5 clusters contain most of the information of magnetic exchange interactions necessary for modeling the magnetic energies revealed by EDM. The case in which ``SQS 1'' is used as testing data shows generally higher testing errors compared with others, because the model has less accuracy for a few sites that have larger deviations in magnetic moments from the average value of $1.834 \mu_\text{B}$, as shown in \Fig{sqsmagmom}. Other ways of data splitting show similar levels of testing RMSEs, suggesting that the Landau-SCE models are well generalized across data from different SQS's.

The same dataset and selected clusters are applied to train the Landau-DNN models and the plot of training and testing RMSEs are presented in the right figure in \Fig{rmses_fccFe}, showing testing RMSEs decreasing to near minima with the first 5 clusters, but switch to increasing with more clusters included. The same ways of data splitting for training and testing datasets as used for the Landau-SCE models are used here. The trends in the testing RMSEs also suggest that the first 5 clusters likely contain most of the information of magnetic exchange interactions necessary for modeling the magnetic energies, similar to the case with Landau-SCE models. The rapid increase in testing errors suggest that large numbers of clusters lead to over-complex neural network models which results in overfitting and poorer generalization of the models. The testing errors for models trained using SQS 1 as testing dataset are not significantly higher than other ways of dataset splitting, suggesting that Landau-DNN models likely handle larger variations in magnitudes of magnetic moments better.

\begin{figure}[htb]
	\makebox[\textwidth][c]{%
		\includegraphics[width=2.2in]{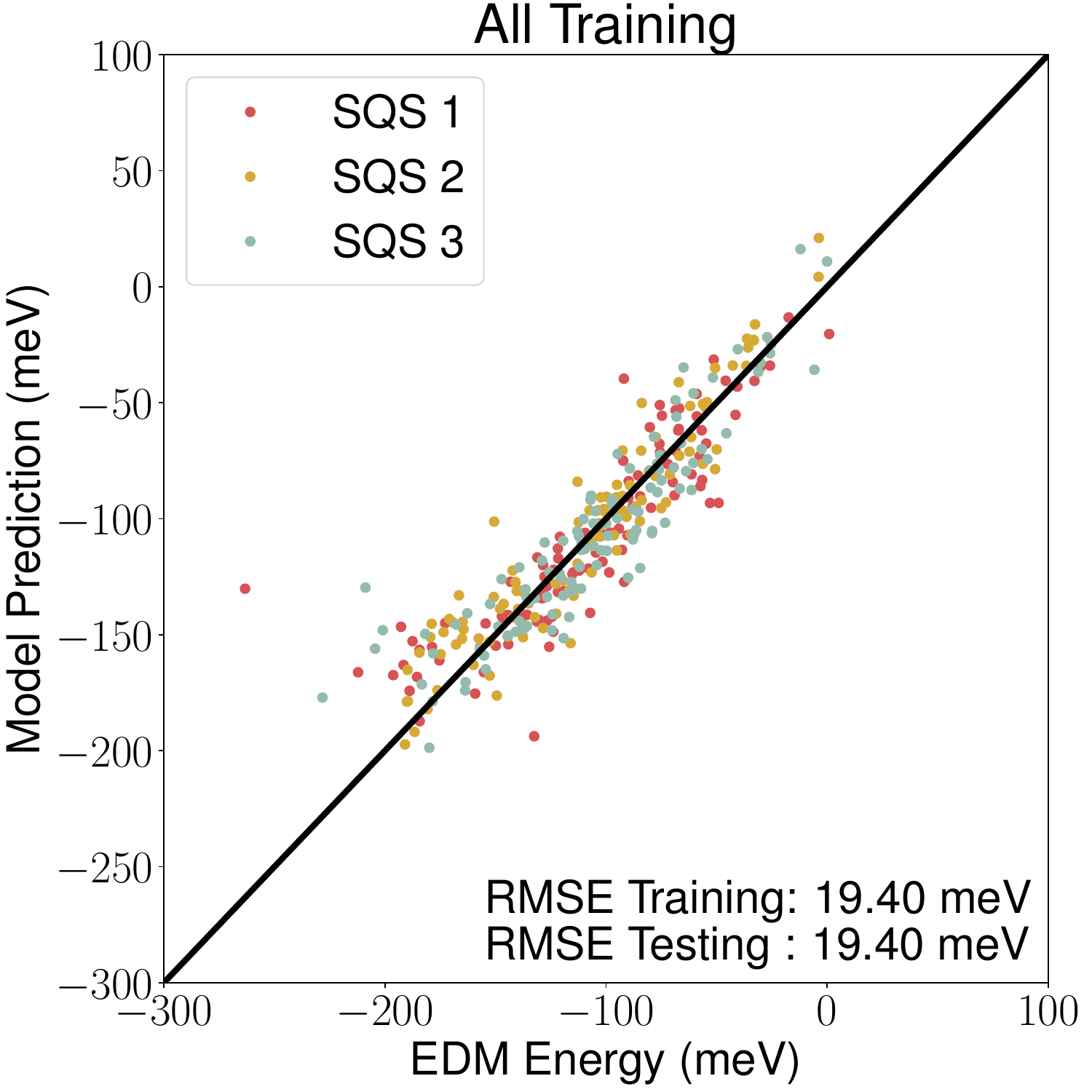} 
		\includegraphics[width=2.2in]{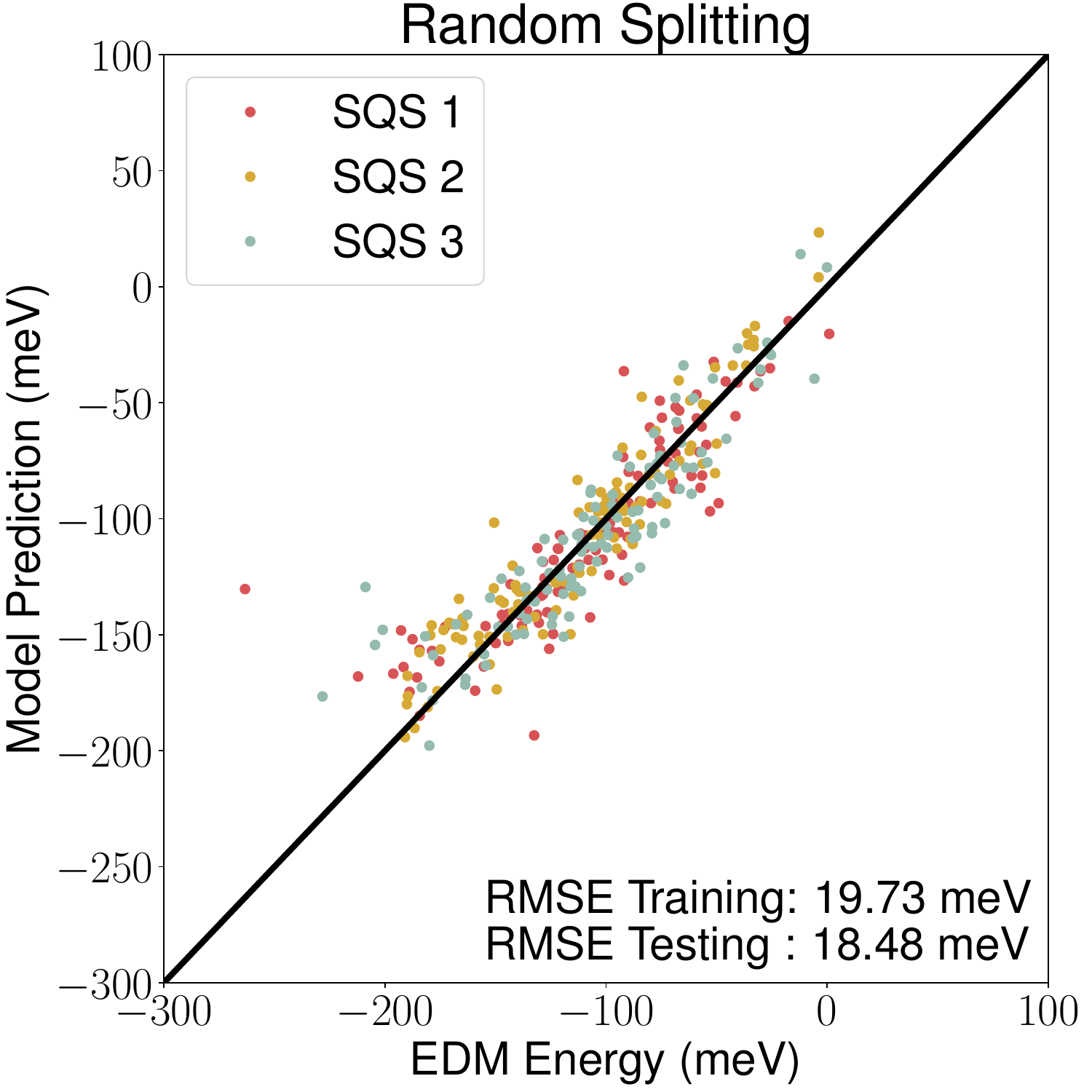}
		\includegraphics[width=2.2in]{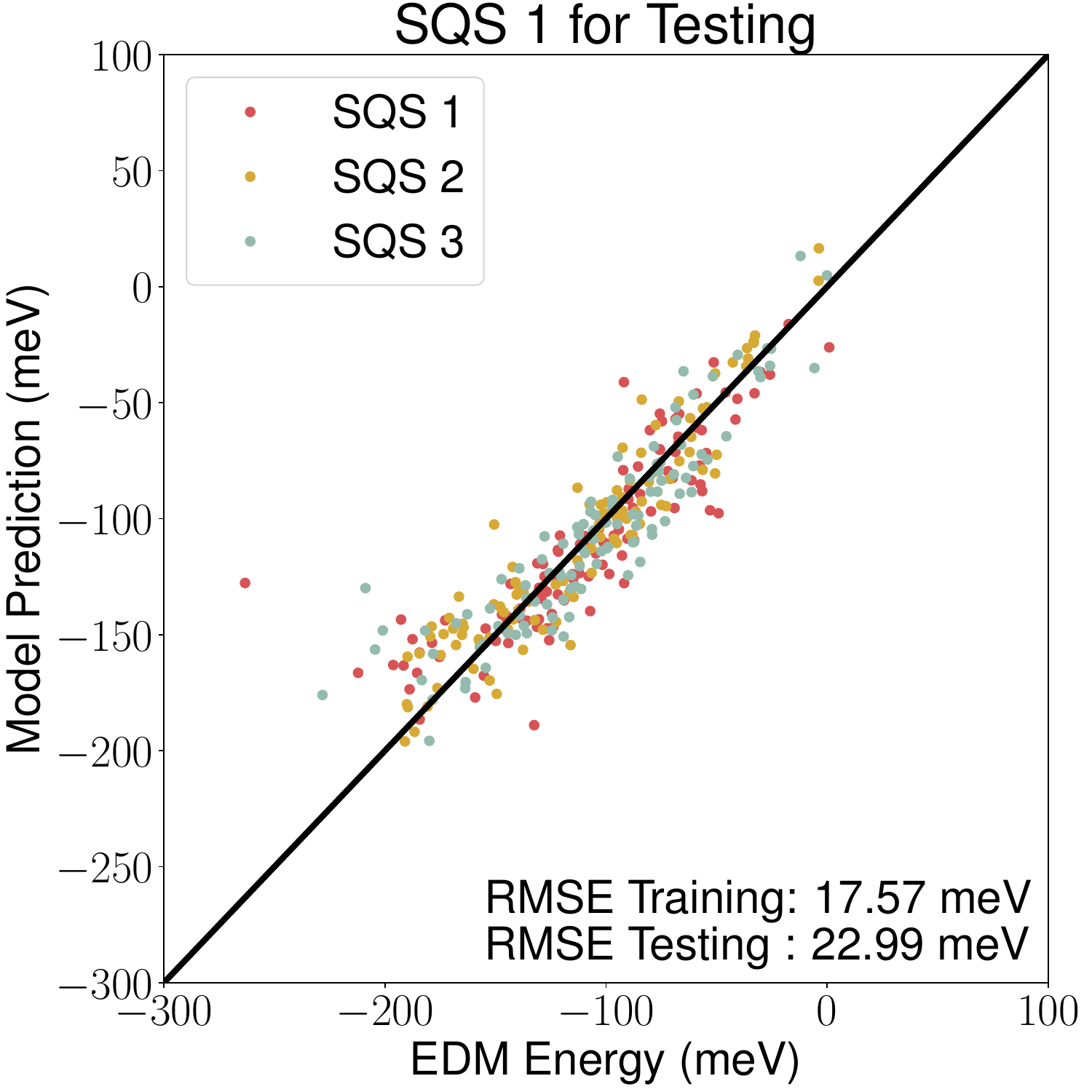}%
	}
	\makebox[\textwidth][c]{%
		\includegraphics[width=2.2in]{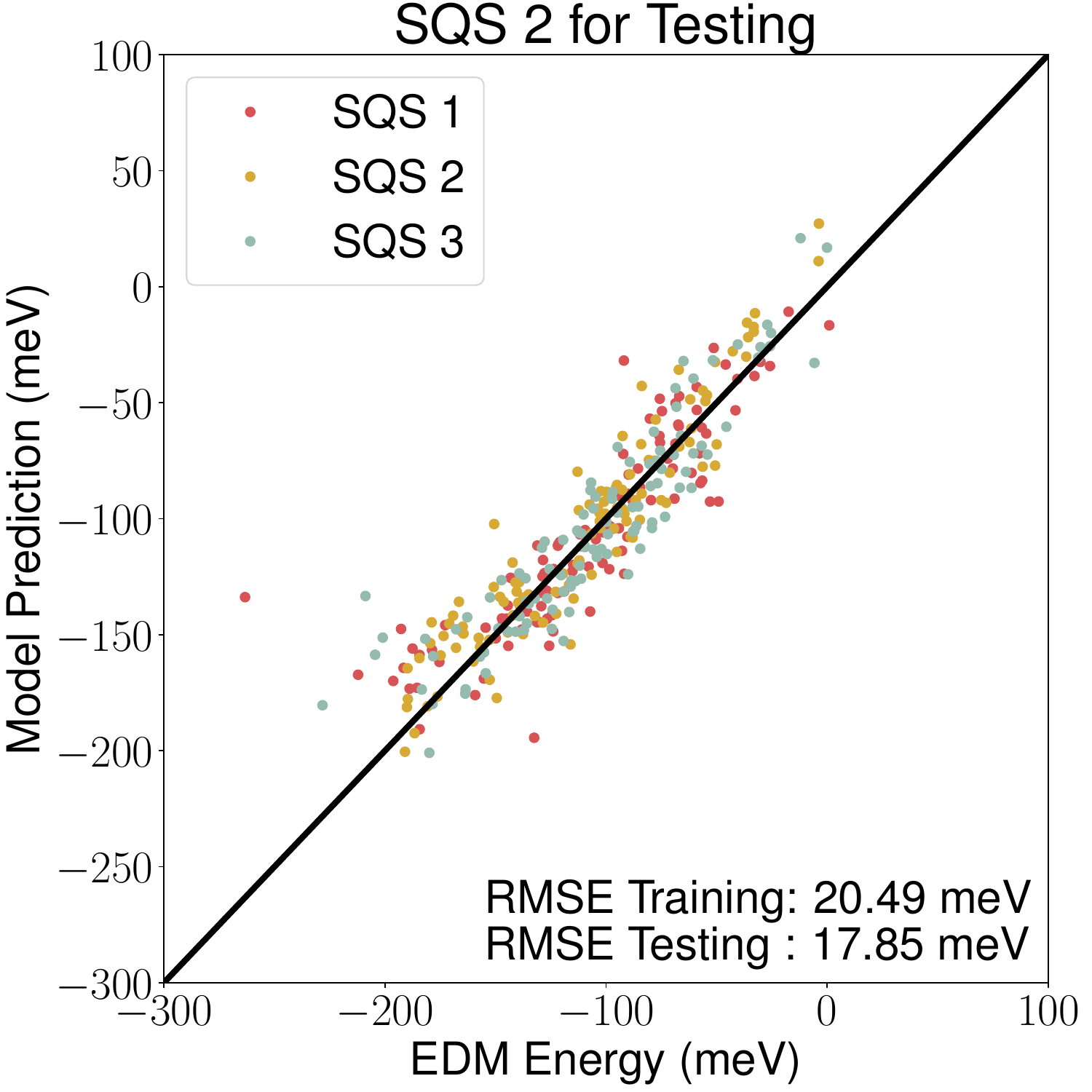}%
		\includegraphics[width=2.2in]{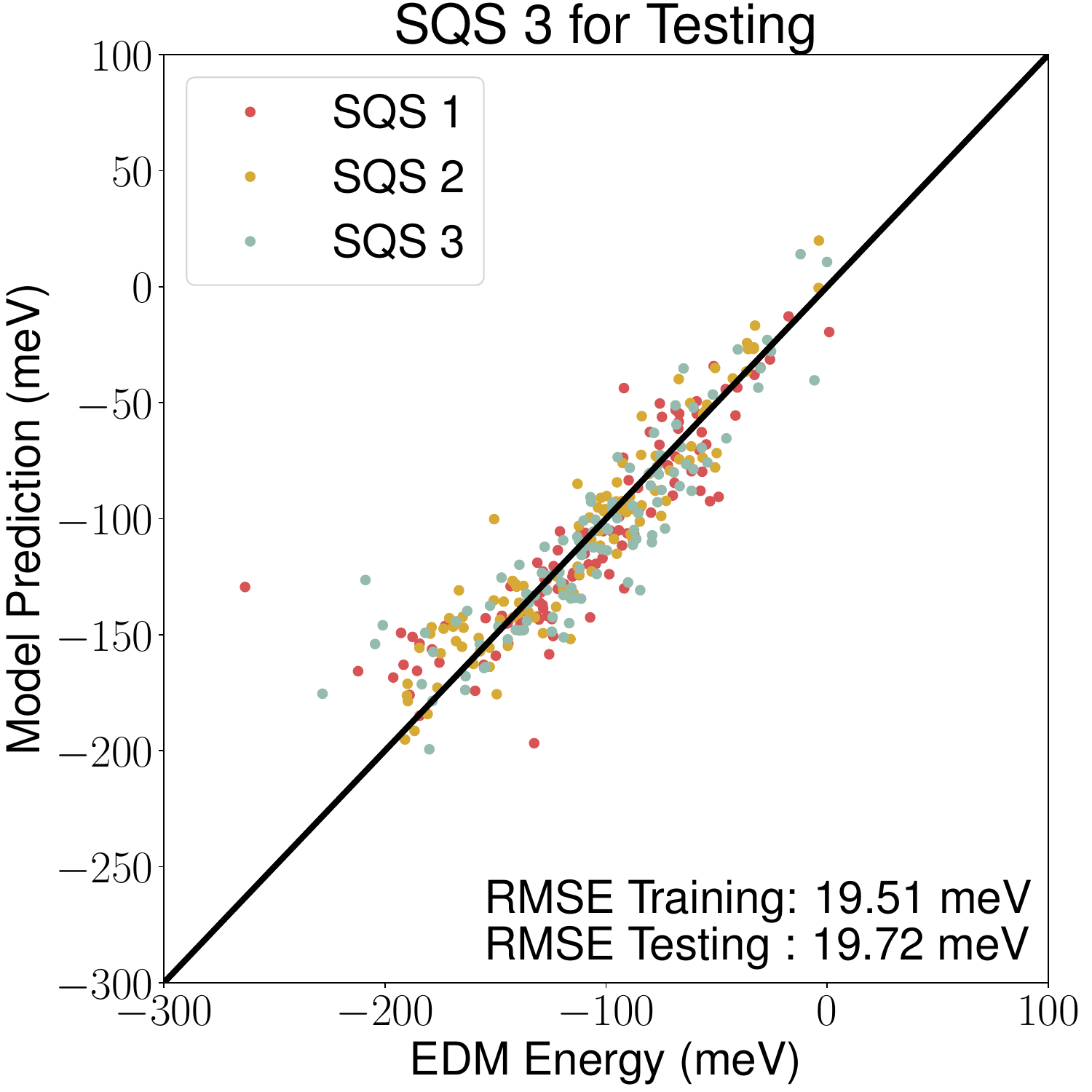}%
	}
	\caption{Energy parity plots for the Landau-SCE models trained using the 5 most important clusters selected by the greedy algorithm. Five ways of training-testing data splitting are performed: use all the data available for training (all training), random 80\%-20\% splitting (random splitting), and SQS-based splitting in which one SQS is taken for testing while the other two SQS's are used for training, with training and testing RMSEs marked in each plot.}
	\label{fig:Eparity_SCE}
\end{figure}

\begin{figure}[htb]
	\makebox[\textwidth][c]{%
		\includegraphics[width=2.2in]{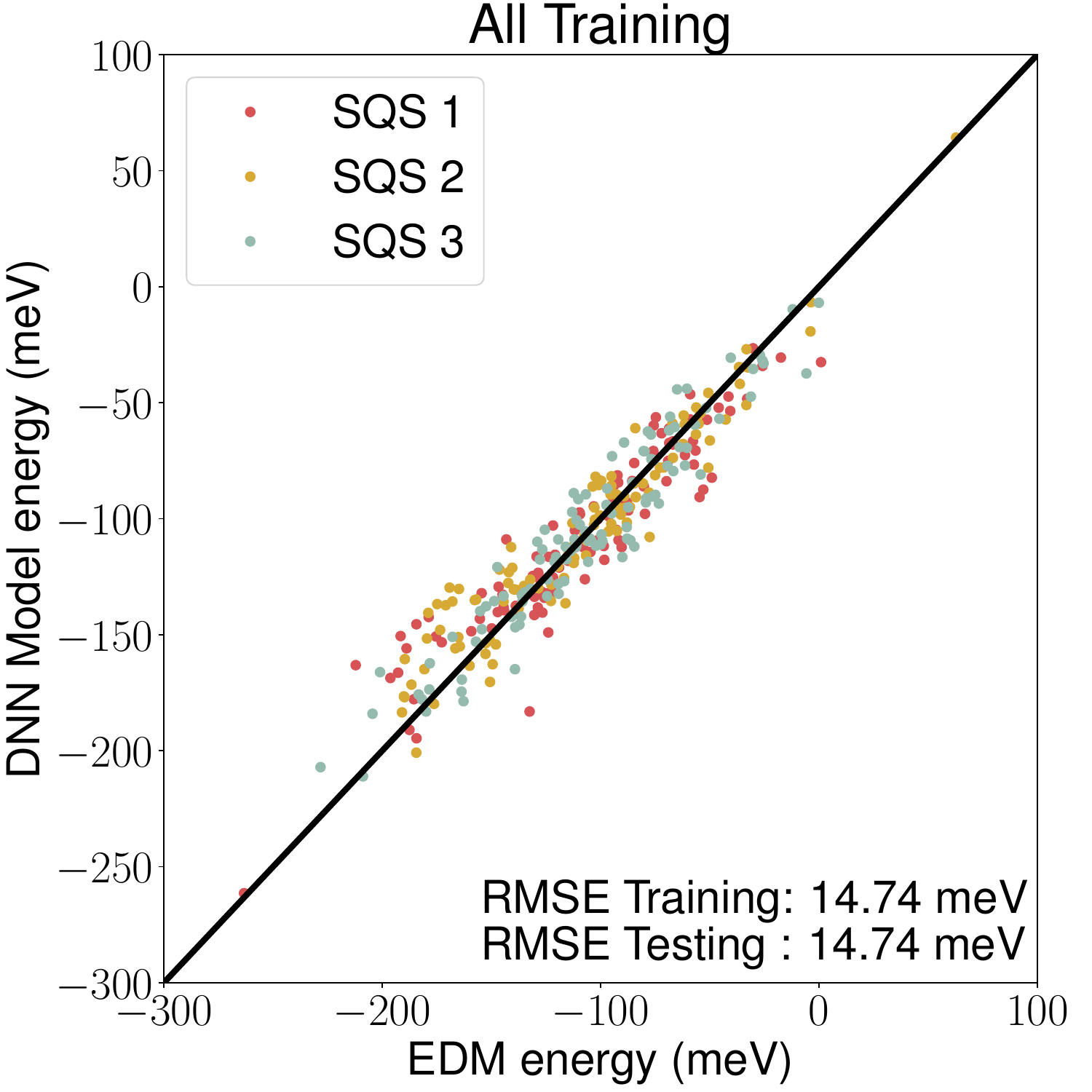} 
		\includegraphics[width=2.2in]{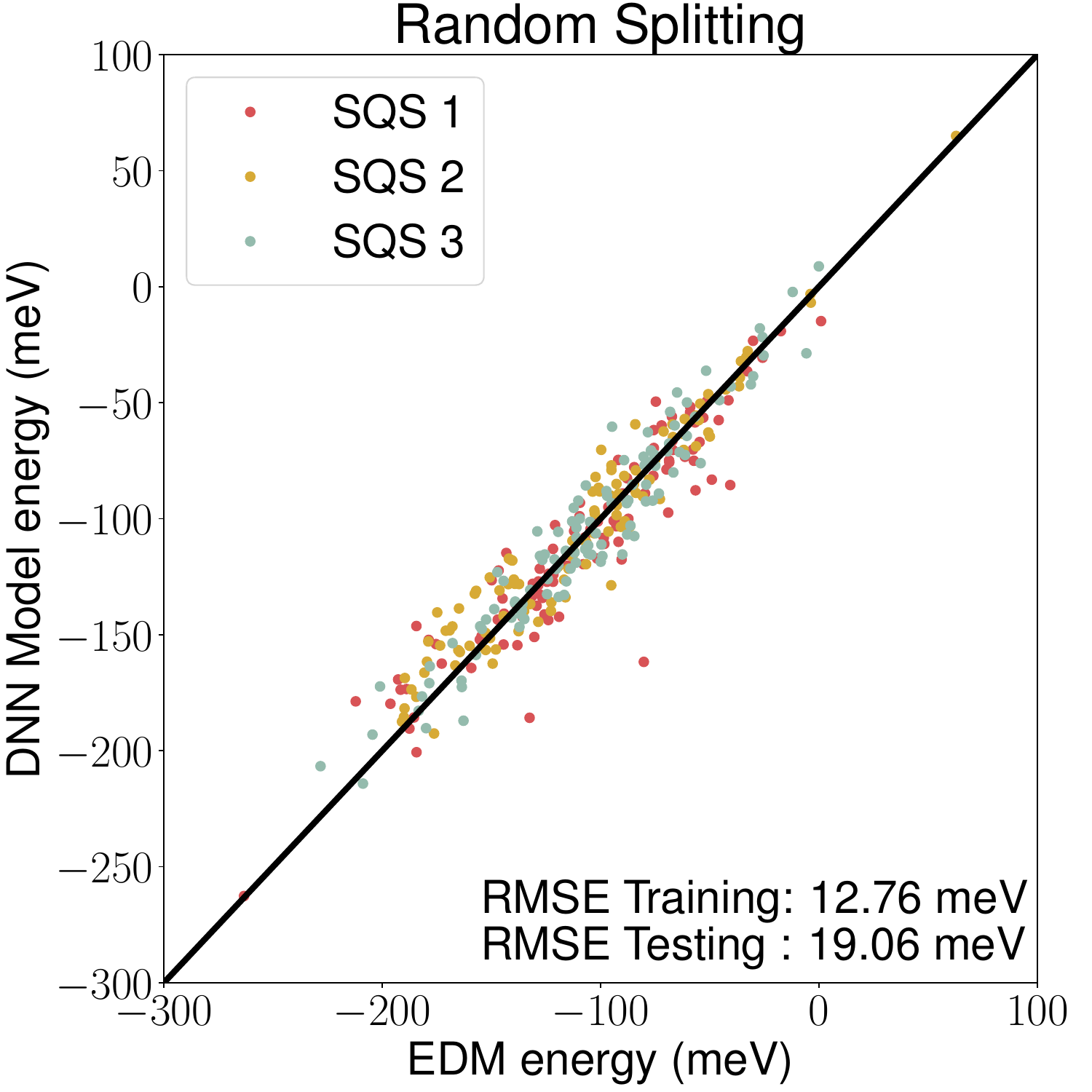}
		\includegraphics[width=2.2in]{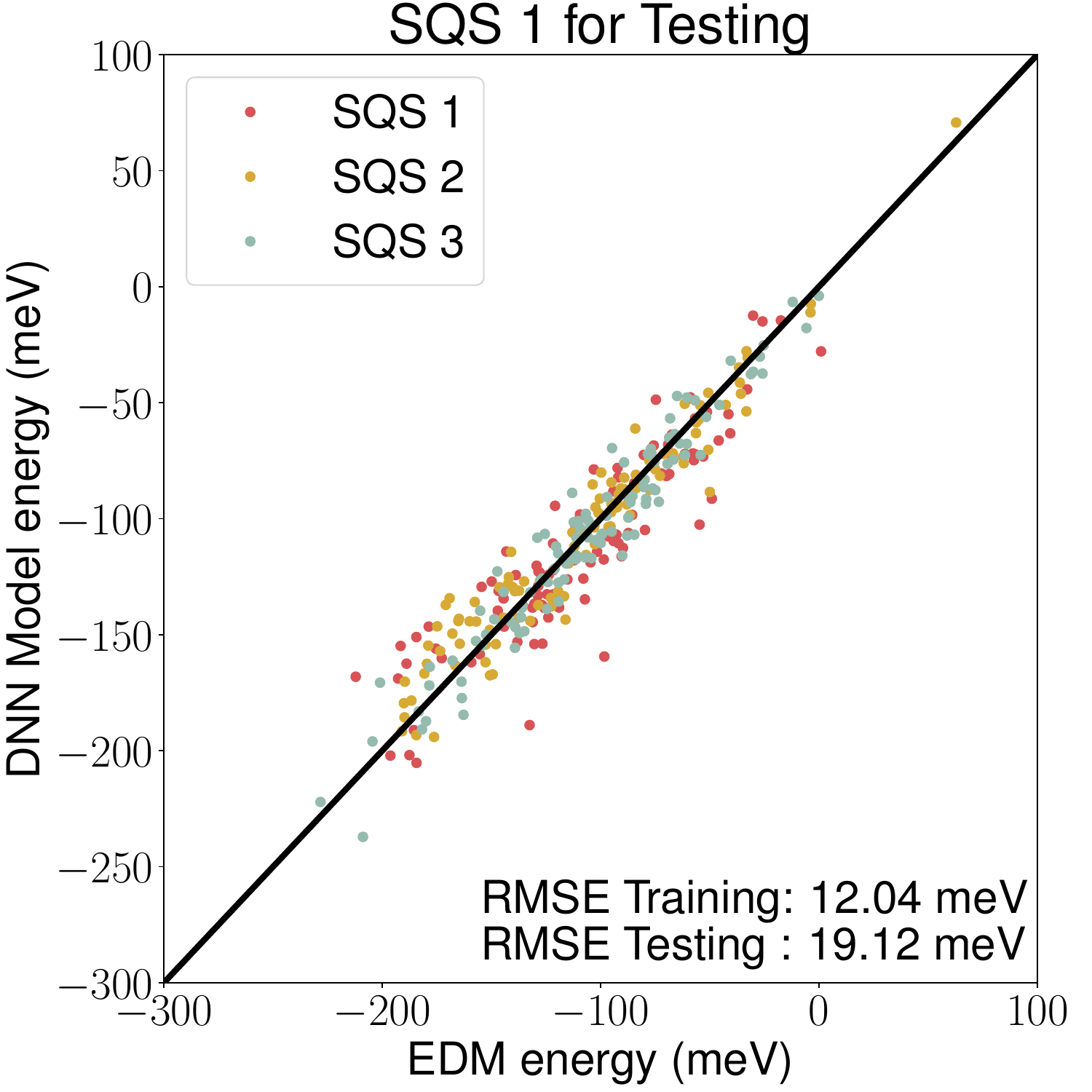}%
	}
	\makebox[\textwidth][c]{%
		\includegraphics[width=2.2in]{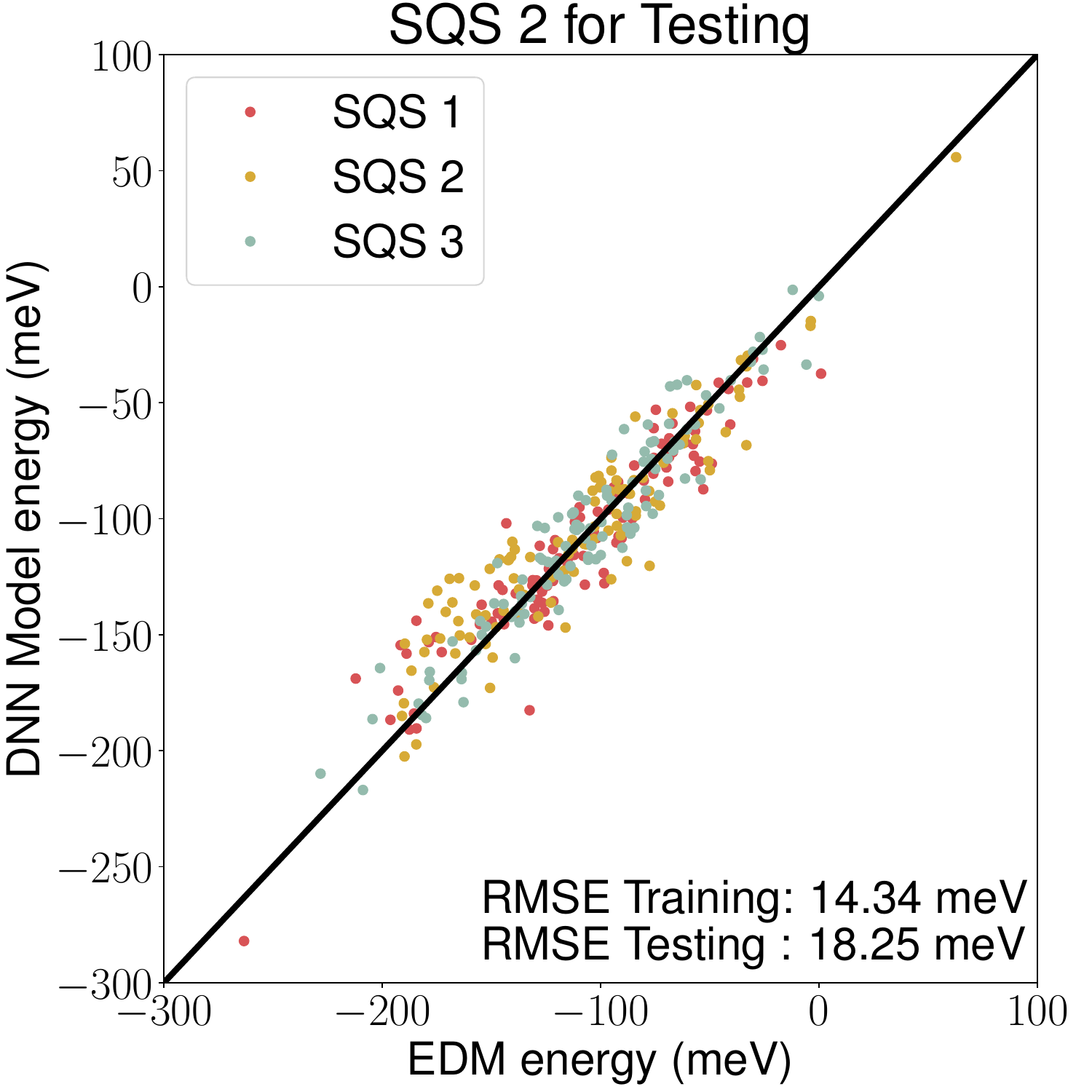}%
		\includegraphics[width=2.2in]{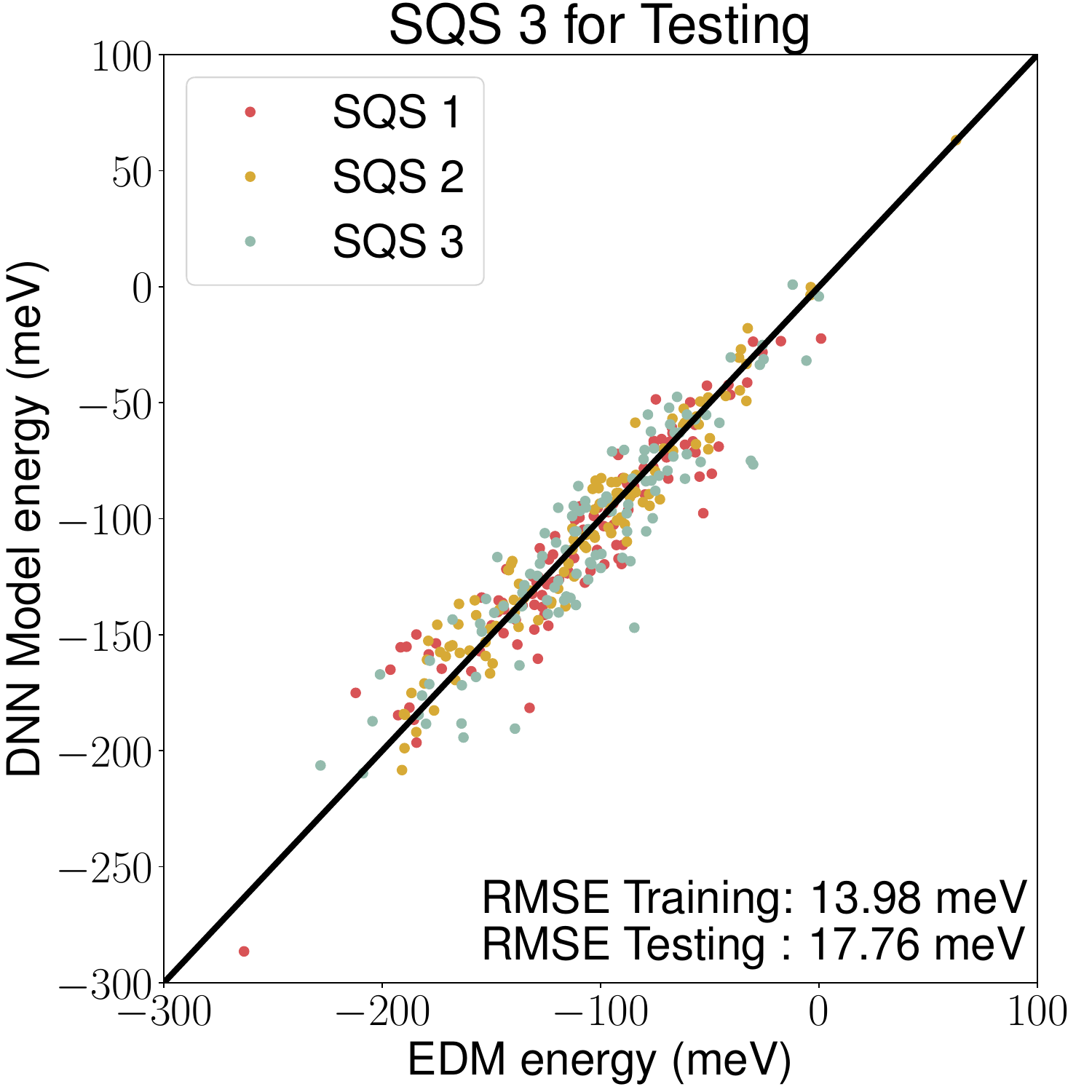}%
	}
	\caption{Energy parity plots for the Landau-DNN models trained using the 5 most important clusters selected by the greedy algorithm. Five ways of training-testing data splitting are performed: use all the data available for training (all training), random 80\%-20\% splitting (random splitting), and SQS-based splitting in which one SQS is taken for testing while the other two SQS's are used for training, with training and testing RMSEs marked in each plot.}
	\label{fig:Eparity_DNN}
\end{figure}

The model RMSE plots in \Fig{rmses_fccFe} indicate that both Landau-SCE and Landau-DNN models created on the first selected 5 clusters (4 pair clusters from the 1st up to the 4th nearest neighbor, and 1 tetrahedron quadruplet cluster) have the overall best balance between accuracy and model complexity; we therefore create our Landau-SCE model and the Landau-DNN model using these clusters, and compare their energy predictions with the ground-truth EDM energies, as presented in \Fig{Eparity_SCE} and \Fig{Eparity_DNN}. Both models show similar accuracies with testing RMSEs being around 20 meV/atom in different ways of training-testing dataset splitting. The parameterized Landau-SCE model can be written as
\begin{equation}
	E_i^\text{Landau-SCE} = -A M_i^2 + B M_i^4 - \sum_{\alpha =1}^4 \sum_{j \in \alpha \text{th N.N.}} J_\alpha^{(2)} M_i M_j - J^{(4)} \sum_{j,k,l} M_i M_j M_k M_l,
\end{equation}
where $J_\alpha^{(2)}$ and $J^{(4)}$ are the coupling constants for the exchange interactions of the pair clusters and the tetrahedron quadruplet cluster, respectively. The parameters of the Landau-SCE model trained with all the three SQS data are listed in \Tab{SCE_coeffs}.

\begin{table*}[htbp]
	\caption{Parameters of the SCE model}
	\centering
	\resizebox{\columnwidth}{!}{
		\begin{tabular}{ccccccc}
			\hline \hline
			 $A$ (meV/$\mu_\text{B}^2$) & $B$ (meV/$\mu_\text{B}^4$) & $J_1^{(2)}$ (meV/$\mu_\text{B}^2$) & $J_2^{(2)}$ (meV/$\mu_\text{B}^2$) & $J_3^{(2)}$ (meV/$\mu_\text{B}^2$) & $J_4^{(2)}$ (meV/$\mu_\text{B}^2$) & $J^{(4)}$ (meV/$\mu_\text{B}^4$) \\
			 \hline
			 $136.1$ & $29.46$ & $-2.312$ & $1.371$ & $0.4316$ & $0.4695$ & $0.008846$ \\
			\hline \hline
		\end{tabular}
	}
	\label{tab:SCE_coeffs}
\end{table*}

The magnetic ordering and exchange interaction parameters obtained in the Landau-SCE model from EDM data align with earlier observations. The DMFT calculations performed on paramagnetic fcc Fe by Katanin \textit{et al.} suggest that the exchange parameters are temperature dependent, with antiferromagnetic ordering for nearest neighbors at low temperatures, and ferromagnetic ordering for nearest and next-nearest exchanges at temperatures higher than the $\alpha-\gamma$ phase transition temperature \cite{Anisimov2018}. They also report exchange parameters of around --6 meV, 6 meV and 1 meV for the first to the third nearest neighbor shells, respectively, at around 386 K with lattice parameter 3.582 \AA; these data are obtained by weight averaging the orbital-resolved interactions with squares of local moments being the weights. Performing a unit conversion by multiplying the mean square magnetic moments in the SQS supercells, EDM gives a close result of $-7.913$, $4.692$ and $1.477$ meV for the first to the third nearest neighbor interactions, respectively. Lavrentiev \textit{et. al.} uses an \textit{ab initio}--parameterized magnetic-cluster-expansion (MCE) model containing Landau self-energies and Heisenberg-type exchange interactions, reporting exchange interaction coefficients $J_1 = -0.650$ $\text{meV}$, $J_2 = 0.267$ $\text{meV}$ and $J_3 = -0.233$ $\text{meV}$ for the first three nearest neighbors, and $A = -65.022$ $\text{meV}$ (their definition of $A$ differ from ours by a sign) and $B = 6.513$ $\text{meV}$ for the self-energy terms \cite{Lavrentiev2010}. This agrees with Katanin \textit{et al.}'s result regarding magnetic ordering for the first two nearest neighbor shells, but at a smaller magnitude; considering that the MCE parameters are fitted using various configurations of pure fcc Fe, including non-magnetic, ferromagnetic and several collinear antiferromagnetic structures, this discrepancy likely comes from the difference in magnitude of magnetic moments in the structures. Singer \textit{et al.} adopted a fully non-collinear SCE model with basis functions dependent on non-collinear spin vectors, reporting $J_1=-17.936$ meV for 1st neighbor interactions and $J_2=12.104$ meV for 2nd neighbor interactions \cite{Singer2011}. 

\subsection{Atomic Energies in Transition Metal-Doped GaN}

Dilute magnetic semiconductors (DMS) are semiconductors doped with a dilute concentration of magnetic ions. The coexistence of semiconducting and magnetic properties, and the convenient manipulation of such properties through doping lead to great potential in various applications, such as spintronics \cite{Sato2010,Pulizzi2012,Dietl2014,Gupta2020,Anbuselvan2021}, magnetic sensors \cite{Punnoose2006}, and quantum dots \cite{Radovanovic2002,Makkar2017}. The magnetic moments of the dopant ions induce spatially localized perturbations to the charge and spin distributions, therefore resulting in perturbations to localized electron energies as well. Here, we apply spin-polarized EDM to quantitatively investigate the perturbations to the atomic energy distributions in GaN, a representative semiconducting material, by dopant Ni ions substituted in the Ga sites. Previous first-principles studies have revealed the connection between the novel electronic, magnetic and optical properties of DMS with spin-dependent effects introduced by the magnetic dopants, such as spin splitting of electronic bands and exchange interactions between localized moments \cite{Katayama-Yoshida2003,Lee2007,Romanudhin2017,Dietl2010}. By enabling the direct calculation of atomic energies, spin-EDM provides a complementary perspective on the interplay between spins and energy-related effects. We first study the case with an isolated magnetic moment, where the self-energy of the magnetic Ni is the only source of perturbation, and subsequently consider exchange interactions with Ni pairs placed at varied separations.

The geometries and electronic structures of the dilute magnetic semiconductor (DMS) Ga$_{1-x}$Ni$_x$N are optimized using DFT, followed by EDM calculations for their atomic energies. Supercells of $3 \times 3 \times 2$ wurtzite GaN are created, with a single Ni atom occupying a Ga site, and two Ni atoms replacing nearest-neighboring and farthest-apart Ga sites, respectively. For the double Ni configurations, two spin configurations are considered: the spin-aligned configuration in which the Ni atoms have the same direction of spins, and the spin-anti-aligned configuration in which the Ni atoms have opposite directions of spins. The Ni spin directions are preserved throughout the optimization process due to weak exchange interactions in the DMS. For pristine GaN, a series of unit cells with varying volumes go through the ionic and cell shape relaxation; their volumes and energies are fitted to the Birch–Murnaghan equation of state for the optimized cell volume and $c/a$ ratio, thus determining the optimized lattice parameters. The ionic positions are then relaxed under the optimized lattice parameters with symmetry of the wurtzite structure enforced. The DFT relaxations are performed using Vienna \textit{Ab initio} Simulation Package (VASP) \cite{Kresse1993,Kresse1994,Kresse1996VASPa,Kresse1996VASPb,Kresse1999PAW} version 5.4.4. Plane-wave basis is used with a cut-off energy 550 eV to converge the DFT total energy to less than 1 meV/atom, with energy tolerance of $10^{-8}$ eV and force tolerance of $1 \text{ meV/\AA}$ for the self-consistency loops. The projector augmented-wave (PAW) pseudopotential \cite{Blochl1994PAW,Kresse1999PAW,Hobbs2000spinPAW} 16 valence electrons ($3p^6 3d^8 4s^2$) for Ni, 13 valence electrons ($3d^{10} 4s^2 4p^1$) for Ga, and 5 valence electrons ($2s^2 2p^3$) for N is used, with Perdew-Burke-Ernzerhof (PBE) functional \cite{Perdew1996GGA} for the exchange-correlation energy. Gaussian smearing with an energy width of $0.03$ eV is applied. The k-point mesh grids are gamma-centered with uniform spacing of around $0.095$ $\text{\AA}^{-1}$ in each dimension in the k-space. The smearing and k-point mesh are chosen so that the density of states (DOS) of bulk GaN near the Fermi level well agrees the DOS calculated by the linear tetrahedron method with Bl\"{o}chl corrections \cite{Blochl1994Tetra} and with a dense k-point mesh, without spin polarization. The initial magnitude of magnetic moments is $2.0 \mu_\text{B}$ for the Ni atoms and $1.5 \mu_\text{B}$ for the Ga and N atoms. The electronic optimization and EDM calculations use the same parameters as the relaxations but with the exception that the density of real-space grids are created with a uniform spacing of around $0.065 \text{ \AA}$ in each dimension, which is twice as dense as used in the DFT relaxations, to ensure that the gauge-dependent errors in EDM atomic energies fall below 1 meV/atom. The atomic energies of bulk GaN from nonspin-polarized EDM calculation and bulk Ni from spin-polarized calculation are used as reference for the EDM energies.

\begin{figure*}[htb]
    \includegraphics[width=0.35\wholefigwidth]{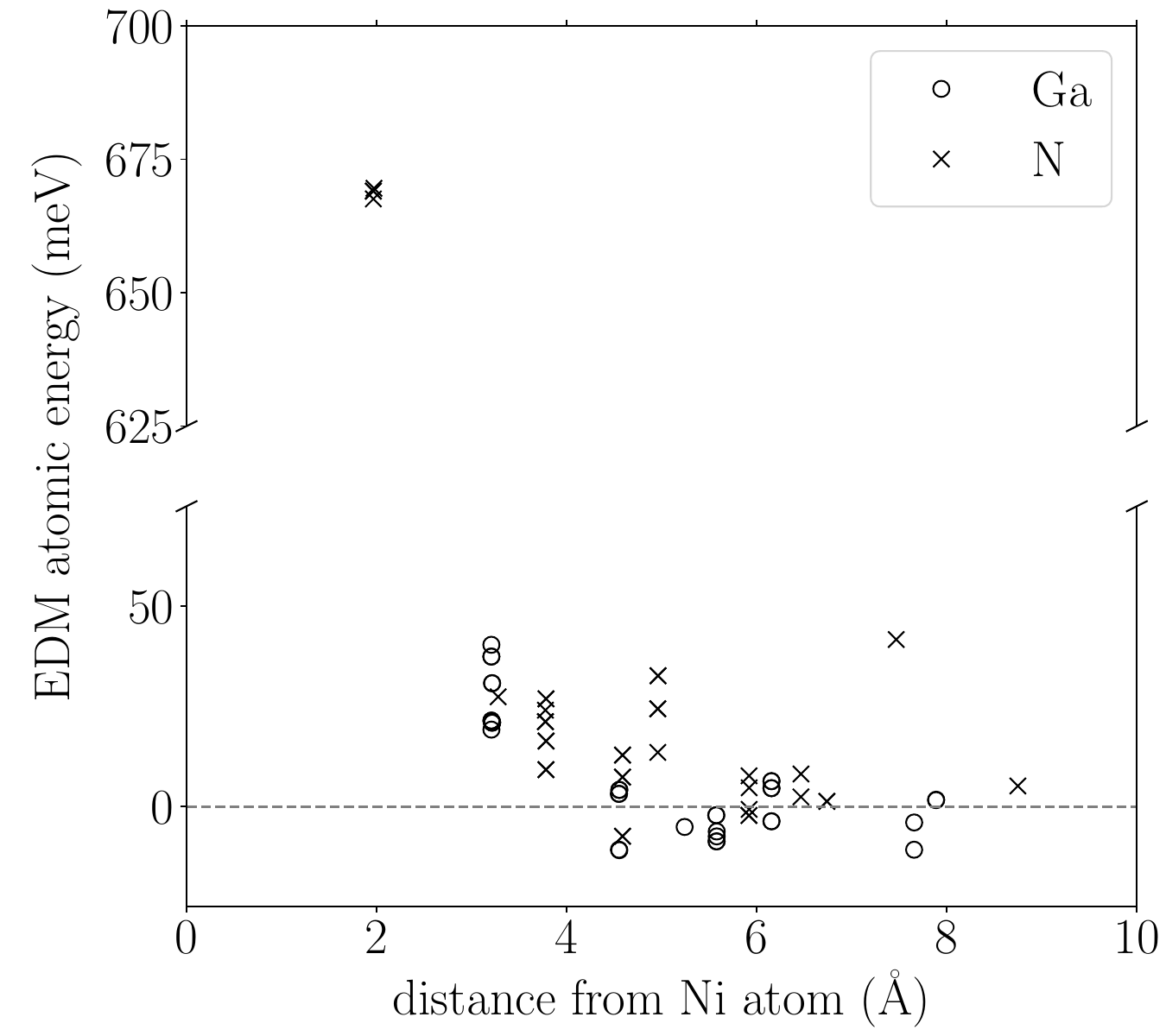}
    \includegraphics[width=0.36\wholefigwidth]{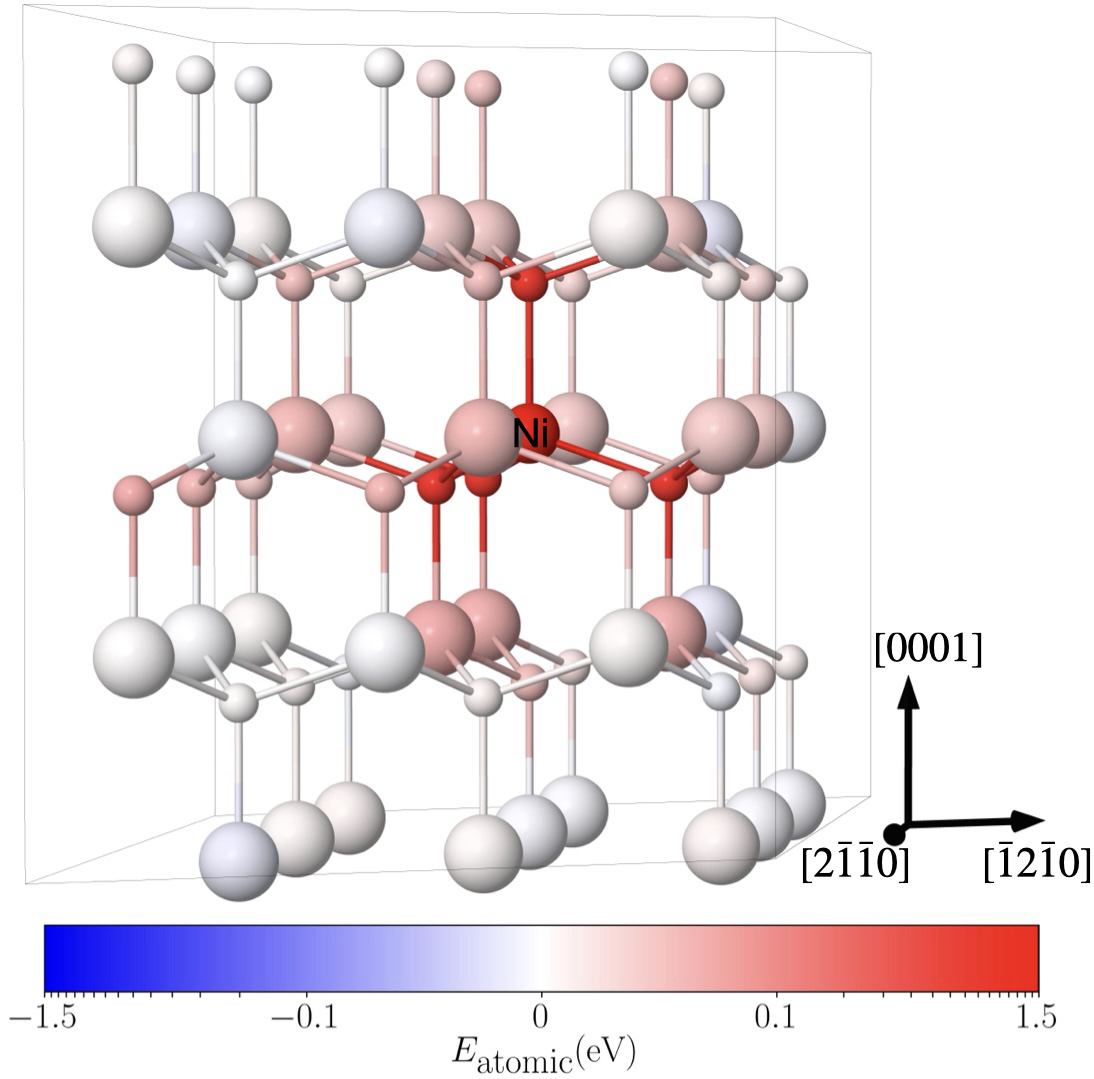}
    \includegraphics[width=0.35\wholefigwidth]{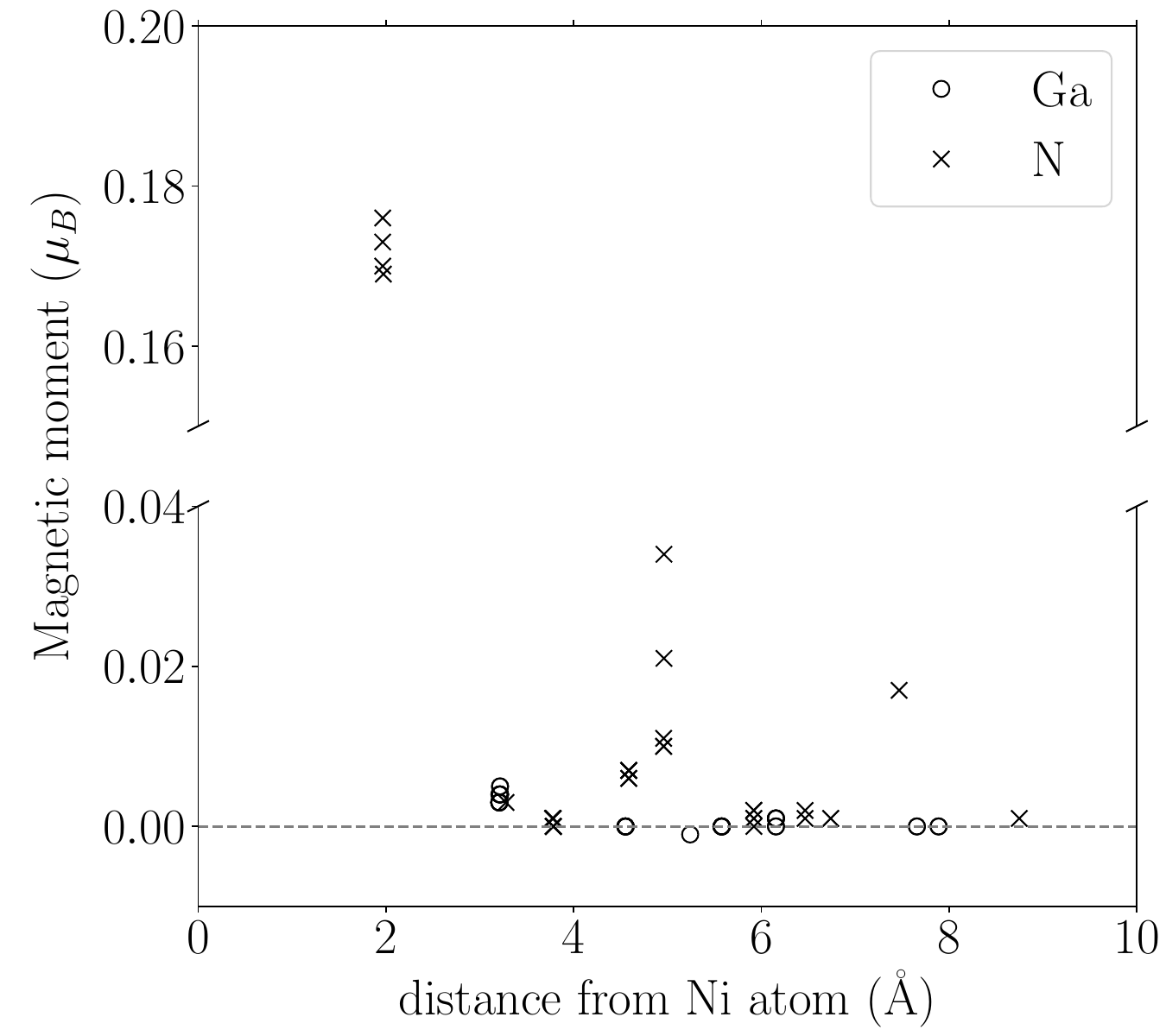}
    
    \caption{EDM atomic energies (left and middle) and magnetic moments of atoms (right) in the $3 \times 3 \times 2$ GaN supercell with a single Ni atom substituting a Ga site (Ga$_{35}$NiN$_{36}$). The EDM atomic energies are plotted with the distance from the substituting Ni atom (left) and as color map in the supercell (middle), while the magnetic moments are plotted with the distance from the Ni (right). Pure GaN with the same lattice constants as Ga$_{35}$NiN$_{36}$ and bulk Ni are used as reference for the EDM energies, which define the zero EDM atomic energies. Energies and magnetic moments are significantly higher for N atoms at the 1st nearest neighbor shell from the Ni atom, and decrease with the distance from the doped Ni atom, suggesting that the perturbation from the substituted Ni atom on the atomic and magnetic structures is strong in the vicinity and decays with distance.}
    \label{fig:1Ni_plots}
\end{figure*}

In the presence of a single substitutional Ni dopant, the magnetic moments and atomic energies of surrounding Ga and N atoms both exhibit deviations from their bulk values that decay with increasing distance from the dopant, as shown in \Fig{1Ni_plots}. The Ni atom is positioned at the center of the supercell and has a magnetic moment of 1.627 $\mu_B$. The four N atoms nearest to the Ni-substituted Ga site, at a distance of 2.0 \AA, show significantly higher energies ($\sim 669\,\text{meV}$) and magnetic moments ($\sim 0.17\, \mu_\text{B}$) than the remaining atoms, indicating strong spin polarization induced by the nearby Ni dopant. Moving away from the Ni, the remaining atoms show drastically smaller energies below 50 meV, and a trend of decaying energies with the distance. A similar trend is seen for magnetic moments as well, which also fall below $0.01 \mu_\text{B}$ for most atoms at distances over 3.0 \AA. The N atoms at 5.0 \AA\ and 7.5 \AA\ from the Ni show slightly increased energies and magnetic moments from the trend due to image interactions: the former is near the face boundary of the periodic supercell and is influenced by another image of the Ni across the boundary, while the latter is near the edge of the supercell and is influenced by three other images of the Ni.

\begin{figure*}[htb]
    \includegraphics[width=1.15\wholefigwidth]{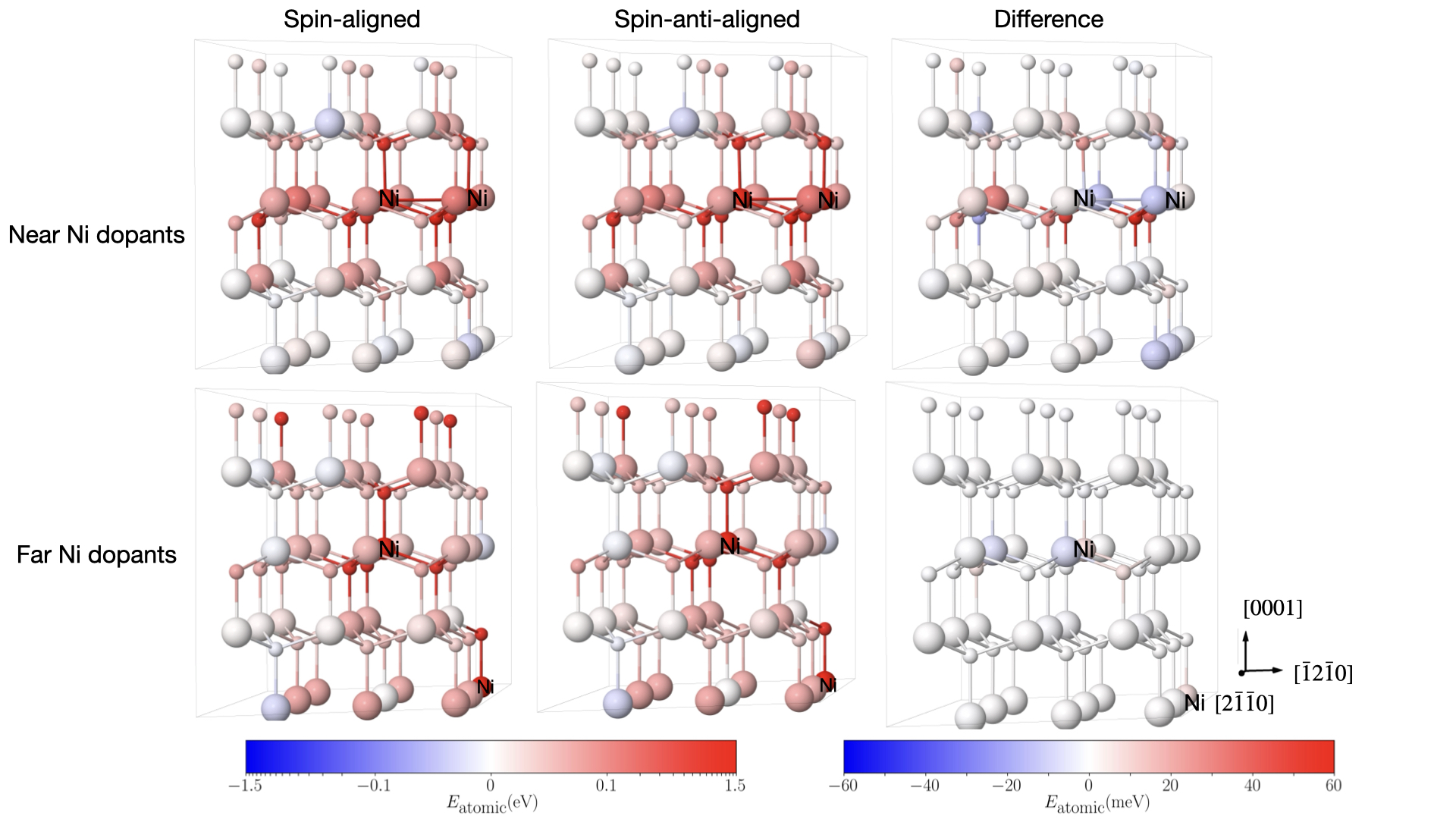}
    
    \caption{Distribution of EDM atomic energies in the $3 \times 3 \times 2$ GaN supercell doped by two Ni atoms (Ga$_{34}$Ni$_{2}$N$_{36}$) at Ga sites, with the two substituting Ni atoms being close as nearest neighbors in the top row, and being far apart in the bottom row. The Ni atoms have aligned spins in the left column, anti-aligned spins in the middle column, and their atomic energy differences shown in the right column. For the nearest neighboring Ni atoms, the magnetic moments are $1.184\, \mu_\text{B}$ and $1.187\, \mu_\text{B}$ for the spin-aligned case, and $\pm 1.434\, \mu_\text{B}$ for the spin-anti-aligned case. For the other case in which the Ni atoms are far apart, the magnetic moments are $1.629\, \mu_\text{B}$ and $1.628\, \mu_\text{B}$ for the spin-aligned configuration, and $-1.624\, \mu_\text{B}$ and $1.623\, \mu_\text{B}$ for the spin-anti-aligned configuration.}
    \label{fig:2Ni_plots}
\end{figure*}

For configurations containing two Ni dopants, the EDM-resolved atomic energy distributions and their differences exhibit a clear dependence on the relative spin alignment and dopant separation, reflecting the influence of exchange interactions between the localized magnetic moments, as shown in \Fig{2Ni_plots}. A total of four configurations are considered: the combination of two atomic configurations in which two Ni atoms are placed at the nearest and farthest Ga sites in the supercell, and two spin configurations in which the Ni atoms have aligned and anti-aligned spins. For each atomic configuration, the EDM energy difference between the two spin configurations is plotted, showing fluctuating atomic energy differences for the near-Ni configuration, while much smaller ($<10 ~\text{meV/atom}$) for the far-Ni configuration. This indicates that the exchange interaction between the Ni atoms, which is strong when the Ni's are near and weak when the Ni's are far, produces different energy perturbations for aligned and anti-aligned spins. The magnetic moments on the Ni atoms in the far-Ni configuration are $1.629\, \mu_B$ and $1.628\, \mu_B$ for aligned spins and $-1.624\, \mu_B$ and $1.623\ \mu_B$ for anti-aligned spins, which are similar in magnitude compared to the $1.627\ \mu_B$ in the case of a single Ni dopant, while in the near-Ni configuration the corresponding values are $1.184\, \mu_B$ and $1.187\, \mu_B$ for aligned spins and $-1.434\ \mu_B$ and $1.434\ \mu_B$ for anti-aligned spins. This shows that the magnetic moments on Ni are suppressed by the exchange interactions. The total EDM energy difference between near-Ni and far-Ni atomic configurations is $-56.8 ~\text{meV}$ for the spin-aligned configuration, and $-320.6 ~\text{meV}$ for the spin-anti-aligned configuration, suggesting that antiferromagnetism is favored by the exchange energy for nearest Ni dopant pairs in GaN. Previous studies revealed the influence of the tetrahedral bonding of Ni/Ga with N on the magnetic ordering of Ni dopants in DMS GaN, with a similar observation that the system is more energetically stable when Ni's are nearest (described as ``forming a dimer with N'' in the original work) \cite{Basha2010}. Opposite spins of Ni atoms in the Ni-N-Ni bond facilitate electron hopping from the shared N with Pauli exclusion principle, leading to more delocalization of electrons compared to the same spins, thus having lower energy.

\section{Conclusion}
In this work, we developed the spin-polarized Energy Density Method (spin-EDM) by extending the Energy Density Method (EDM) to the framework of spin-density functional theory for magnetic systems. Spin-EDM computes localized atomic energies by integrating spatial energy density functions over gauge-invariant volumes. The DFT total energy is decomposed into a real-space energy density function, incorporating spin-dependent kinetic, Coulomb, exchange-correlation, and on-site terms. This methodology allows for efficient extraction of localized atomic energies influenced by their specific magnetic environments. We applied spin-EDM to two systems: paramagnetic fcc Fe, and Ni-doped dilute magnetic semiconductor (DMS) GaN. For fcc Fe, Landau-SCE and Landau-DNN models are fitted to the EDM energies and DFT magnetic moments calculated from three SQS structures, which predict atomic energies from atomic descriptors constructed from SCE. With five clusters selected by the greedy algorithm, the models are able to predict atomic energies with RMSE below 20 meV/atom. For Ni-doped GaN, spin-EDM reveals distance-decaying localized energy perturbations for the single Ni configuration, and exchange interactions that favor antiferromagnetic states between nearby dopants for the double Ni configuration. Future work could extend the formalism by relaxing the current assumptions of collinearity and neglected spin-orbit coupling, enabling the study of more complex magnetic phenomena such as non-collinear spin textures. Additionally, integrating spin-EDM datasets with advanced machine learning architectures offers a promising pathway for developing highly accurate magnetic interatomic potentials, which could ultimately accelerate the design of novel spintronic devices and advanced magnetic alloys.

\section*{Acknowlegements}
This work is funded by the U.S. National Science Foundation under grant number NSF/MPS-1940303. The research is performed using computational resources provided by the Stampede 2 supercomputer of the Texas Advanced Computing Center (TACC) at The University of Texas at Austin, and by the Illinois Campus Cluster, which is operated by the Illinois Campus Cluster Program (ICCP) in conjunction with the National Center for Supercomputing Applications (NCSA) and supported by funds from the University of Illinois at Urbana-Champaign. The geometry and atomic energy distribution figures are generated using Jmol \footnote{Jmol: an open-source Java viewer for chemical structures in 3D. \href{http://www.jmol.org/}{http://www.jmol.org/}}.

\section*{Code and Data Availability}
The EDM code is available on GitHub \cite{EDMGitHub}. The data is available at the Materials Data Facility \cite{Blaszik2016, Blaszik2019}, doi:10.18126/hj3n-0g80 \cite{Yang2026FeData} and doi:10.18126/jx65-kk33 \cite{Yang2026GaNData}.


%
\end{document}